\documentclass[aps,prl,twocolumn,superscriptaddress,longbibliography,10pt,colorlinks]{revtex4-2}

\usepackage[english]{babel}
\usepackage[utf8]{inputenc}

\usepackage{amsthm}
\usepackage{amsmath,bm,bbm}
\usepackage{amssymb}
\usepackage{amsfonts}
\usepackage{graphicx}
\graphicspath{{manuscript_figures/}}
\graphicspath{{./figures/}}
\usepackage{txfonts} \usepackage{xcolor}
\usepackage{float}
\usepackage{braket}
\usepackage{booktabs}
\usepackage{comment}
\usepackage{ulem}
\usepackage{orcidlink}

\usepackage{makecell}
\usepackage{multirow}

\usepackage{hyperref}

\usepackage{bbold}
\usepackage{siunitx}
\makeatletter
\let\arxiv@mkpream\@mkpream
\makeatother

\newcommand{\avg}[1]{\langle#1\rangle}

 \newcommand{\abs}[1]{\left\vert #1 \right\vert}

\newcommand{\proj}[1]{\ket{#1}\bra{#1}} \newcommand{\braopket}[3]{\langle #1 | #2 | #3\rangle} \newcommand{\evSmall}[1]{\langle #1 \rangle} 
\newcommand{\trSmall}[1]{\text{Tr}[#1]}

\newcommand{\evTextSub}[2]{\text{E}_{#1}\left[#2\right]}

\newcommand{\varSub}[2]{\text{Var}_{#1}\left[#2\right]}

     \newcommand{\deriv}[2]{\frac{\text{d}#1}{\text{d}#2}}
\newcommand{\IQ}{\mathcal{I}_Q}
\newcommand{\IC}{\mathcal{I}_C}

\allowdisplaybreaks

\usepackage[framemethod=tikz]{mdframed}
\newmdenv[
  leftmargin=10pt,
  rightmargin=0pt,
  innerleftmargin=10pt,
  innertopmargin=0pt,
  innerbottommargin=0pt,
  skipabove=\topsep,
  skipbelow=\topsep,
  linecolor=gray,
  linewidth=0.5pt,
  frametitleaboveskip=0pt,
  frametitlebelowskip=0pt,
  rightline=false,
  topline=false,
  bottomline=false
]{quotebar}

\newtheorem{claim}{\textbf{Claim}}

\begin{document}

\title{Stochastic signal sensing with finite energy and dead time at the fundamental quantum limit}

\author{James W. Gardner\,\orcidlink{0000-0002-8592-1452}}
\email{jamesgardner@uchicago.edu}
\affiliation{Chicago Quantum Institute, Pritzker School of Molecular Engineering, University of Chicago, Illinois 60637, USA}

\author{Tuvia Gefen\,\orcidlink{0000-0002-3235-4917}}
\email{tuvia.gefen@mail.huji.ac.il}
\affiliation{Racah Institute of Physics, The Hebrew University of Jerusalem, Jerusalem 91904, Givat Ram, Israel}

\author{Matteo Fadel\,\orcidlink{0000-0003-3653-0030}\,}
\email{fadelm@phys.ethz.ch}
\affiliation{Department of Physics, ETH Z\"{u}rich, 8093 Z\"{u}rich, Switzerland}

\date{\today}

\begin{abstract}
    State preparation, measurement, and reset operations take finite time and use finite energy in realistic experiments, yet the impact of this on optimal quantum metrological protocols is not properly understood. We study the effect on sensing a stochastic signal, relevant for the detection of ultralight dark matter and other searches for fundamental physics. We prove that two-mode squeezed vacuum is the optimal probe state given a finite mean-energy constraint for a family of incoherent sensing problems, including noise sensing and quantum illumination. For estimating a gain independent of a loss, we show that entanglement is a required resource to achieve the fundamental quantum limit and observe a non-Gaussian to Gaussian transition in the optimal unentangled state as the dead time increases. We apply our results to bulk acoustic wave resonators.
\end{abstract}

\maketitle

\begin{figure}[t]
    \centering
    \includegraphics[width=\columnwidth]{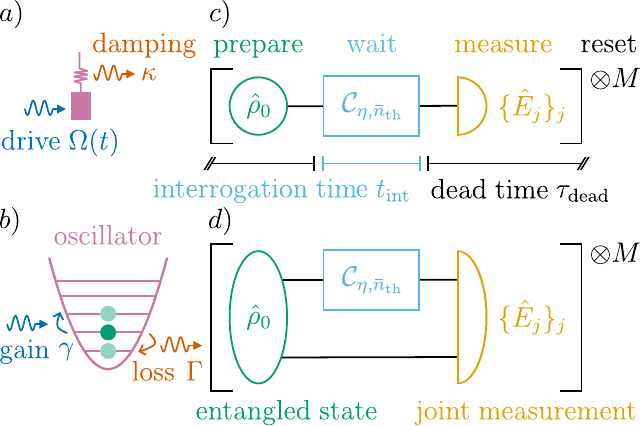}
    \caption{\textbf{a)} Driven, damped mechanical oscillator with drive $\Omega(t)$ and damping $\kappa$. \textbf{b)} Quantum harmonic oscillator with gain $\gamma$ and loss $\Gamma$.     \textbf{c--d)} Prepare-wait-measure-reset metrology, where each of the $M$ cycles takes an interrogation time $t_\text{int}$ plus a dead time $\tau_\text{dead}$ from the combined state preparation, measurement, and reset operations. We show both the \textbf{(c)} unentangled and \textbf{(d)} entangled strategies.}     \label{fig:diagram dead time}
\end{figure}

Sensing a weak stochastic signal is an important task in quantum metrology. In the continuous-variable setting, stochastic signals model searches for ultralight dark matter~\cite{lamoreaux2013analysis,dixit2021searching,shi2023ultimate,shi2025quantum}, stochastic gravitational waves~\cite{gardner2025stochastic,renzini2022stochastic}, and geontropic quantum gravity~\cite{mcculler2022single,vermeulen2025photon}. These problems are related more broadly to Lindblad estimation, where the task is to estimate the weak rate of a known Lindblad jump operator~\cite{gardner2025lindblad,sekatski2022optimal,das2025universal}. Noise spectroscopy of a linear system in this way~\cite{tsang2012fundamental,ng2016spectrum} is also connected to quantum superresolution for optical imaging~\cite{tsang2023quantum,tsang2016quantum,nair2016far,oh2021quantum,PhysRevLett.127.130502,gardner2026algorithmic}, as well as being the stochastic analogue of deterministic waveform estimation~\cite{Tsang+2011,PhysRevLett.132.130801,ding2026holevo}. In spin systems, stochastic signals can model Pauli Lindblad estimation problems such as qubit spectroscopy~\cite{mouradian2021quantum} and nanoscale nuclear magnetic resonance~\cite{gefen2019overcoming,cohen2020achieving}. The fundamental limits of stochastic signal sensing have been studied previously~\cite{shi2023ultimate,gardner2025stochastic,gardner2025lindblad,shi2025quantum,sekatski2022optimal,das2025universal,gorecki2025interplay}, but thus far with the unrealistic assumption that state preparation, measurement, and reset operations take negligible time in the experiment. This leads to the optimal scheme being a continuous quantum jump experiment~\cite{nagourney1986shelved}. 

In this work, we find bounds and optimal protocols for a generic class of bosonic stochastic signals, accounting for realistic limitations such as finite energy and dead time.
In particular, the thermal loss channel, i.e.\ a beamsplitter with a thermal state, is a ubiquitous model of a stochastic signal arising in thermometry~\cite{sekatski2022optimal}, noise sensing for dark matter searches~\cite{shi2023ultimate}, and quantum illumination and radar~\cite{PhysRevLett.118.070803}.
Our main result is that two-mode squeezed vacuum is optimal for estimation of any parameter of the thermal loss channel given finite energy. 
The practical limitation that state preparation, measurement, and reset operations take a finite dead time to perform affects the time-optimized signal-to-noise ratio.
We generalize the bounds and protocols of Ref.~\cite{gorecki2025interplay} to account for these realistic limitations of finite energy and dead time.
We show that these restrictions change the optimal protocols and their behavior, e.g.\ the energy dependence of the fundamental bound, in a nontrivial manner.

\vspace{2mm}
\textbf{Stochastic signals.}--- Let us demonstrate how the thermal loss channel arises naturally. Consider sensing the amplitude of a weak signal using a resonant bosonic mode, as shown in Fig.~\ref{fig:diagram dead time}. 
In the rotating frame, the master equation is thus: 
\begin{align}\label{eq:Hamiltonian}
    \deriv{\hat \rho}{t} = -\frac{i}{\hbar} [\hat H, \hat\rho] + \kappa \mathcal{D}[\hat a]\hat \rho, \quad
    \hat H=\hbar \left(\Omega(t) \hat a^\dagger + \Omega^\ast(t) \hat a \right),
\end{align}
where $\hat a$ is the annihilation operator satisfying $[\hat a, \hat a^\dagger]=1$, $\Omega(t)$ is the complex amplitude of the drive of interest, $\kappa$ is the damping rate, and $\mathcal{D}[\hat a]\hat \rho = \hat a\hat \rho\hat  a^\dagger - \frac{1}{2}\{\hat a^\dagger \hat a,\hat \rho\}$.

The case of a coherent, phase-locked signal is well known. In this case, $\Omega(t)=\Omega_0 e^{i\phi}$ is time-independent with $\Omega_0$ to be estimated and $\phi$ known. 
Given an interrogation time $t_\text{int}$, state evolution is described by the Gaussian channel $\mathcal{E}_{t_\text{int},\phi}=\mathcal{T}_{\alpha}\circ\mathcal{L}_{\eta}$, which is a vacuum loss channel $\mathcal{L}_\eta$ with transmissivity $\eta=e^{-\kappa t_\text{int}}$ followed by a displacement channel $\mathcal{T}_\alpha$ with complex displacement $\alpha=-i\frac{2\Omega_0}{\kappa}(1-\sqrt{\eta})e^{i\phi}$~\cite{fadel_quantum_2025}. 
In comparison, the thermal loss channel $\mathcal{C}_{\eta,\bar n_\text{th}}$ corresponds to a beamsplitter with transmissivity $\eta$ with a thermal state of average particle number $\bar n_\text{th}$. This acts as follows on a single-mode Gaussian state with mean vector $\vec\mu\to\sqrt{\eta}\vec\mu$ and covariance matrix $\Sigma\to\eta\Sigma+(1-\eta)\Sigma_\text{th}$, where $\Sigma_\text{th}=(\frac{1}{2}+\bar n_\text{th})\,\text{diag}(1,1)$. This is the solution to the master equation:
\begin{align}\label{eq:me}
    \deriv{\hat \rho}{t} = \gamma \mathcal{D}[\hat a^\dagger]\hat \rho + \Gamma \mathcal{D}[\hat a]\hat \rho, \quad \Gamma:=\kappa+\gamma,
\end{align}
where $\bar n_\text{th}=\gamma/\kappa$.
The thermal loss channel represents a couple of different models of stochastic signals.
The first model is a coherent signal probed by a sensor that is not phase locked with the source, i.e.\ $\Omega(t)=\Omega_0 e^{i\phi}$ but with $\phi$ unknown. Marginalizing the phase $\phi$ uniformly over $(0, 2\pi)$, the resulting channel $\bar{\mathcal{E}}_{t_\text{int}}=\frac{1}{2\pi}\int_0^{2\pi}\text{d}\phi\;\mathcal{E}_{t_\text{int},\phi}$ is non-Gaussian in general since it maps, e.g., the vacuum to a mixed ring of coherent states.
For weak signals $\abs{\alpha}^2(\bar N+1)\ll1$ with $\alpha$ defined above, however, it is approximately the Gaussian thermal loss channel $\mathcal{C}_{\eta,\bar n_\text{th}}$ with $\eta=e^{-\kappa t_\text{int}}$ and $\bar n_\text{th}=\frac{4\Omega_0^2}{\kappa^2}\tanh\left(\frac{\kappa t_\text{int}}{4}\right)$ for any given $t_\text{int}$~\cite{supplemental}. 
In the short time limit $\kappa t_\text{int}\ll1$, this Gaussianity condition then becomes $\Omega_0^2t_\text{int}^2(\bar N+1)\ll1$ as $\abs{\alpha}\approx\Omega_0t_\text{int}$ and $\bar n_\text{th}\approx\frac{\Omega_0^2t_\text{int}}{\kappa}$. 
The second model of a stochastic signal is a complex-valued Gaussian process $\Omega(t)$, e.g.\ Lorentzian noise with correlation time $\tau_\phi$ and the following correlators:
\begin{align}\label{eq:stochavg}
\avg{\Omega(t)} = \avg{\Omega(t) \Omega(t')} =0, \quad \avg{\Omega(t) \Omega^*(t')}=\Omega_0^2 e^{-|t-t'|/\tau_\phi}.
\end{align} 
In the limit of short correlation time $\tau_\phi\ll t_\text{int}$, this signal results in diffusion at a rate of $\gamma=2\Omega_{0}^2 \tau_\phi$. Combined with the loss rate $\kappa$ from Eq.~\ref{eq:Hamiltonian} for $\kappa \tau_\phi\ll 1$ recovers the master equation in Eq.~\ref{eq:me} and the thermal loss channel, as required.

In summary, the thermal loss channel $\mathcal{C}_{\eta,\bar n_\text{th}}$ arises naturally as an effective model of a stochastic signal, where information about the signal $\Omega(t)$ is encoded in $\bar n_\text{th}$ or equivalently $\gamma$ for a given $\kappa$~\cite{shi2023ultimate}. Separately, we consider estimating $\gamma$ for a fixed $\Gamma$, which surprisingly behaves differently even if $\gamma\ll\kappa$. Moreover, this channel arises also in quantum illumination, radar, and reading tasks, in which the problem is to estimate $\eta$ for a given $\bar n_\text{th}$~\cite{PhysRevLett.118.070803, jonsson2022gaussian}. Therefore, we consider the broad task of channel estimation of a generic parameter of $\mathcal{C}_{\eta,\bar n_\text{th}}$. We will tackle this after first reviewing estimation theory, dead time, and the coherent signal case.

\vspace{2mm}
\textbf{Quantum parameter estimation.---} We use the quantum Fisher information (QFI) to determine the limit of estimating some parameter $\theta$. In amplitude units, the signal-to-noise ratio is $\text{SNR}=\theta/\Delta\theta$, where $\Delta\theta$ is the standard deviation of an unbiased estimator of $\theta$.
We use the SNR because it is dimensionless and qualitatively similar for different choices of the parameter, e.g.\ the SNRs with respect to $\theta$ and $\theta^2$ are a factor of two apart. (Hence, we estimate $\sqrt\gamma$ or $\sqrt\kappa$ later instead of $\gamma$ or $\kappa$ to avoid a coordinate singularity in the QFI that does not affect the SNR.)
The classical Cram\'er--Rao bound states that $\text{SNR}\leq\text{SNR}_C:=\theta\sqrt{M\IC}$, where $M$ is the number of independent and identically distributed repetitions of the experiment and $\IC=\sum_j(\partial_\theta p_j)^2/p_j$ is the classical Fisher information (CFI) of the probability distribution of measurement results $p_j=\trSmall{\hat \rho(t_\text{int})\hat E_j}$, where $\{\hat E_j\}_j$ is a set of effects describing the measurement \cite{kay1993statistical}. This bound is asymptotically tight for $M\gg1$ and achieved by maximum likelihood estimation. 

Maximizing the CFI over the measurement scheme, the quantum Cram\'er--Rao bound states that $\text{SNR}\leq\text{SNR}_Q:=\theta\sqrt{M\IQ}$, where $\IQ=\text{Tr}[\hat \rho(t_\text{int})\hat L^2]$ is the quantum Fisher information (QFI) and $\hat L$ solves the Lyapunov equation $\partial_\theta\hat \rho(t_\text{int})=\{\hat \rho(t_\text{int}),\hat L\}/2$ \cite{braunstein1994statistical}. The eigenbasis of $\hat L$ constitutes an optimal measurement such that $\text{SNR}_C=\text{SNR}_Q$. The Lyapunov equation can be solved numerically using the spectral decomposition of $\hat \rho(t_\text{int})$. 

We now maximize the QFI over the initial state $\hat\rho_0$. By convexity of the QFI, the initial state is pure and the measurement projective without loss of generality. We impose a constraint on the average particle number $\langle\hat n\rangle=\trSmall{\hat\rho_0\hat a^\dagger\hat a}\leq\bar N$. Depending on whether we allow the initial state to be entangled with a noiseless ancilla, we define the entanglement-assisted channel QFI (ECQFI) as $\IQ^\text{E}:=\max_{|\Psi\rangle\in\mathcal{H}\otimes \mathcal{H}_{A},\langle\hat n\rangle=\bar{N}}\IQ$ where $\rho_0=\proj{\Psi}$ and the unentangled channel QFI (UCQFI) as $\IQ^\text{U}:=\max_{|\psi\rangle\in\mathcal{H},\langle\hat{n}\rangle=\bar{N}}\IQ$ where $\hat \rho_0=\proj{\psi}$.
These imply bounds $\text{SNR}_Q^j$ on the SNR for $j\in\{\text{E},\text{U}\}$. While $\IQ^\text{E}\geq \IQ^\text{U}$ always holds, there is an entanglement gap if $\IQ^\text{E}> \IQ^\text{U}$ such that entanglement is required to achieve the ECQFI.

\vspace{2mm}
\textbf{Dead time.---} State preparation, measurement, and reset operations take finite dead time $\tau_\text{dead}$ in any realistic experiment. As a first approximation, we assume that the dead time for a given $\bar N$ and $\kappa$ is independent of the chosen state and measurement. For the prepare-wait-measure-reset strategy shown in Fig.~\ref{fig:diagram dead time}, the number of repetitions is $M\approx T/(t_\text{int}+\tau_\text{dead})$ given a total experiment time of $T$. In the asymptotic large $T$ limit, the SNR is thus:
\begin{align}\label{eq:SNR}
    \text{SNR}\leq\text{SNR}_Q^j\approx\theta\sqrt{\frac{\IQ^jT}{t_\text{int}+\tau_\text{dead}}},\quad j\in\{\text{E},\text{U}\}.
\end{align}
The standard quantum limit–scaling $\text{SNR}\propto\sqrt{T}$ is expected. We maximize this bound over the interrogation time $t_\text{int}$, for a given dead time $\tau_\text{dead}$ and average particle number $\bar N$, to define the time-optimized (TOP) SNR as:
\begin{align}\label{eq:TOP}
    \text{SNR}_Q^{j,\text{TOP}}:=\max_{t_\text{int}}\text{SNR}_Q^j, \quad j\in\{\text{E},\text{U}\},
\end{align}
where we label the optimal interrogation time as $t_\text{int}^*$. There is a tradeoff between the number of cycles $M$ and the QFI per cycle $\IQ^j$. 
This has been studied in, e.g., Refs.~\cite{dooley_quantum_2016,chu2023strong, herb2024quantum}, but we extend it to stochastic signals.

\vspace{2mm}
\textbf{Coherent signal sensing.---} Let us first review the precision limits of the canonical displacement estimation problem, where we estimate $\Omega_0$ from the coherent, phase-locked signal $\mathcal{E}_{t_\text{int},\phi}$. 
There is no entanglement gap as a single-mode squeezed vacuum state with $\bar N=\sinh^2(r)$ and quadrature measurement is optimal, leading to the following ECQFI~\cite{latune2013quantum,demkowicz2013fundamental}.
\begin{align}\label{eq:coherent signal, SMSV}
    \IQ^\text{E} = \dfrac{16 \left(1-\sqrt{\eta}\right)^{2}}{\kappa^2\left[1 - \eta\left(1-e^{-2r}\right)\right]}. 
\end{align} In the high-energy limit, this is $(16/\kappa^2) \tanh(\kappa t_\text{int}/4)$ which is bounded with $\bar N$. 
                    In the high-energy limit $t_\text{int}^*\gg e^{-2r}/\kappa$, the TOP SNR is $(\text{SNR}_Q^{\text{E},\text{TOP}})^2=\frac{4\Omega_0^2T}{\kappa}\text{sech}^2(\kappa  t_\text{int}^*/4)$. 
For zero dead time, the TOP SNR squared is $4\Omega_0^2T/\kappa$, which is the fundamental limit when optimized over all possible quantum control and requires fast measurements~\cite{demkowicz2017adaptive,zhou2018achieving,gorecki2025interplay}.
For short dead times $e^{-6r}/\kappa\ll \tau_\text{dead}\ll 1/\kappa$, then this is $[1-(3\kappa\tau_\text{dead})^{2/3}/4]\times 4\Omega_0^2T/\kappa$ for $t_\text{int}^*\approx 2(3\tau_\text{dead}/\kappa^2)^{1/3}$. Whereas, for long dead times $\kappa \tau_\text{dead}\gg1$, then it is $16\Omega_0^2T/(\kappa^2 \tau_\text{dead})$ for $t_\text{int}^*\approx \frac{2}{\kappa}\log(\kappa \tau_\text{dead})$.
Hence, the dead time changes the precision bounds and $t_\text{int}^{*}$, but it does not change the optimal input state or the required resources in this case. We want to see if this is different in the stochastic case.

\vspace{2mm}
\textbf{Thermal loss channel.---} Let us now estimate a generic parameter $\theta$ of the thermal loss channel $\mathcal{C}_{\eta,\bar n_\text{th}}$, where $\eta=e^{-\kappa t_\text{int}}$ and $\bar n_\text{th}=\gamma/\kappa$, which represents a family of incoherent sensing problems. We start by proving the following result:
\begin{claim}\label{claim:ECQFI and TMSV}
    Consider sensing a parameter $\theta$ of the single-mode Gaussian thermal loss channel $\mathcal{C}_{\eta,\bar n_\text{th}}$ with transmissivity $\eta$ and bath occupation $\bar n_\text{th}$, both implicitly functions of $\theta$. The initial state is constrained by $\evSmall{\hat a^\dagger\hat a}\leq\bar N$ given an average particle number of the probe $\bar N$. The ECQFI with respect to $\theta$ is achieved by preparing a two-mode squeezed vacuum (TMSV) state with a noiseless ancilla mode and is given as follows:
    \begin{align}\label{eq:ECQFI}
        \IQ^\text{E}=\frac{c_1\bar N^2+c_2\bar N+c_3}{c_4\bar N+c_5},
    \end{align}
    where the verbose coefficients $c_n$, given in the End Matter, are independent of $\bar N$. \end{claim}
We prove this claim in the Supplemental Material~\cite{supplemental} by calculating a purification-based upper bound on the ECQFI~\cite{escher2011general} and showing that the TMSV QFI attains it~\cite{monras2013phase}. This result holds even for finite $\bar N$ (relevant for experiments) and finite $\theta$ and $t_\text{int}$, thus generalizing many previous results that assumed specific encoding or fast measurements~\cite{PhysRevLett.118.070803,sekatski2022optimal,jonsson2022gaussian, shi2023ultimate,gardner2025stochastic,gardner2025lindblad, gorecki2025interplay, brady2026precision}.
Preparing TMSV with a low-noise ancilla can pose additional experimental challenges and lower the tolerance to undesired noise~\cite{gardner2025stochastic}. We therefore want to also understand the performance of states without entanglement~\cite{wang2024exponential}.

There are many ways to decompose and parameterize the thermal loss channel~\cite{holevo2007one}. We discuss the ECQFI and TOP SNR for the following cases: $\theta=\sqrt{\gamma}$ for fixed $\kappa$, $\theta=\sqrt\kappa$ for fixed $\bar n_\text{th}$, and $\theta=\sqrt\gamma$ for fixed $\Gamma$. These cases show that the scaling with $\bar N$ and entanglement gap can change dramatically. 

\vspace{2mm}
\textbf{Noise sensing.---} Let us estimate $\theta=\sqrt{\gamma}$ (equivalently, $\bar n_\text{th}$) for a fixed $\kappa$, such that Eq.~\ref{eq:ECQFI} then becomes:
\begin{align}\label{eq:noise sensing}     \IQ^\text{E}=\frac{4(1-\eta)[(2\gamma+\kappa)\bar N+\gamma+\kappa]}{(\gamma+\kappa)[(1-\eta)\{(2\gamma+\kappa)\bar N+\gamma\}+\kappa]}.
\end{align}
This establishes that the TMSV QFI from Ref.~\cite{shi2023ultimate} is optimal for all $\bar N$. The high-energy limit of $4/(\gamma+\kappa)$ is bounded and there is no entanglement gap as there exist unentangled non-Gaussian protocols that are also optimal~\cite{gardner2025stochastic}. The simultaneous weak-signal ($\gamma\ll\kappa$) and high-energy ($\kappa \tau_\text{dead}\ll\bar N$) limit of the TOP SNR is~\cite{supplemental}: 
\begin{align}\label{eq:TOP noise sensing}
    (\text{SNR}_Q^{\text{E},\text{TOP}})^2 \approx \frac{4 \gamma T (\bar N+1)}{\left(1+\sqrt{\kappa\bar N \tau_\text{dead}}\,\right)^2},
\end{align}
where $t^*_\text{int}=\sqrt{\tau_\text{dead}/(\kappa\bar N)}$ scales differently than for the coherent signal. 

For short dead times $\kappa \tau_\text{dead} \ll \min(1/\bar N ,\bar N )$, the TOP SNR squared in Eq.~\ref{eq:TOP noise sensing} is $\left(1-2\sqrt{\kappa\bar N\tau_\text{dead}}\right)\times4 \gamma T (\bar N+1)$ which grows linearly with $\bar N$ while the condition holds. 
This short dead time limit recovers the result that the optimal protocol is a continuous quantum jump experiment~\cite{sekatski2022optimal,gardner2025lindblad}. Here, any pure state satisfying $\evSmall{\hat a}=\evSmall{\hat a^2}=0$ is optimal, e.g.\ TMSV or any Fock state or superposition of Fock states spaced by at least three particles~\cite{gardner2025lindblad, gorecki2025interplay}.
Let us consider zero dead time, such that we may assume that $\kappa t_\text{int}\ll1$. Then, the thermal loss channel approximation to the first stochastic signal model is valid provided that $\Omega_0^2t_\text{int}^2(\bar N+1)\ll1$. This implies that in the zero dead time limit, the TOP SNR squared of $\bar{\mathcal{E}}_{t_\text{int}}$, given by $4\gamma T(\bar N+1),$ is worse than the TOP SNR squared of the coherent case, $\mathcal{E}_{t_\text{int},\phi}$, which is $4\Omega_0^2T/\kappa$. This is expected, since in the stochastic case of $\bar{\mathcal{E}}_{t_\text{int}}$ we lose the information about $\phi$ compared to the coherent case of $\mathcal{E}_{t_\text{int},\phi}$. 

For longer dead times, we observe a transition in the TOP SNR squared in Eq.~\ref{eq:TOP noise sensing}. For $1/\bar N \ll\kappa  \tau_\text{dead} \ll \bar N$, it becomes instead $4\gamma T/(\kappa \tau_\text{dead})$ for $\bar N\gg1$ which is bounded in $\bar N$ for finite dead time. This means that the dead time significantly changes the behavior of the precision limit, from unbounded to bounded.

We study the isotropic thermal loss channel here. For the asymmetric thermal loss channel from Ref.~\cite{gardner2025stochastic}, while TMSV is optimal in the high energy limit, it is not always optimal at finite energy as, e.g., a single-mode squeezed state can exploit the asymmetry to perform up to a factor of two better~\cite{supplemental}.

\vspace{2mm}
\textbf{Loss estimation.---} We now estimate $\theta=\sqrt\kappa$ (equivalently, $\eta$) for a fixed $\bar n_\text{th}$. This variously represents quantum illumination to detect a remote target, estimating the relaxation time of a mechanical oscillator, or sensing the loss per unit length of an optical fiber. The ECQFI in Eq.~\ref{eq:ECQFI} is:
\begin{align}\label{eq:loss estimation}
    \IQ^\text{E} = \frac{4\kappa\eta t_\text{int}^2[(1-\eta ) \bar N^2+\left(2 \eta  \bar n_{\text{th}}+1\right)\bar N +\eta  \bar n_{\text{th}}]}{(1-\eta)^2 \left(2 \bar n_{\text{th}}+1\right)\bar N +(1-\eta) \left[(1-\eta) \bar n_{\text{th}}+1\right]}.
\end{align}
This establishes that TMSV is optimal for all $\bar N$ and any $\eta$ and $\bar n_\text{th}$, in comparison to the loose bound in the quantum illumination limit of $\eta\to0$ and $\eta\bar N\ll(1-\eta)\bar n_\text{th}$ from Ref.~\cite{PhysRevLett.118.070803}.
In the high energy limit of $(1-\eta) \bar N\gg2 \bar n_{\text{th}}+1$, the ECQFI is $4\kappa\eta t_\text{int}^2\bar N/[(1-\eta)\left(2 \bar n_{\text{th}}+1\right)]$ which is unbounded and grows linearly with $\bar N$, unlike for noise sensing, because $c_1\propto\dot \eta^2$. There is an entanglement gap of at most a factor of two to the UCQFI, which is achieved by a coherent state for $\eta\to0$~\cite{PhysRevLett.118.070803}. Using the Gaussian QFI~\cite{monras2013phase}, the coherent state QFI is:
\begin{align}\label{eq:coherent QFI, illumination}
    \IQ = \frac{4\kappa\eta t_\text{int}^2\bar N}{2 (1-\eta) \bar n_\text{th}+1}+\frac{4\kappa\eta^2 t_\text{int}^2\bar n_\text{th}}{(1-\eta) [(1-\eta) \bar n_\text{th} + 1]},
\end{align}
such that the entanglement gap closes for $\bar N\gg1$ and $\eta\to0$, but the coherent state QFI does not attain the ECQFI for finite~$\eta$. 
For short dead times, $[(2 \bar n_{\text{th}}+1)/\bar N]^2/2\ll \kappa \tau_\text{dead}\ll1$, the optimal interrogation time is $t^*_\text{int}\approx\sqrt{2\tau_\text{dead}/\kappa}$ which scales similarly to noise sensing. The TOP SNR is thus~\cite{supplemental}:
\begin{align}\label{eq:illumination, harsh approx}
    (\text{SNR}_Q^{\text{E},\text{TOP}})^2 \approx \left(1-\sqrt{2\kappa\tau_\text{dead}}\right)\times\frac{4\kappa T\bar N}{2 \bar n_{\text{th}}+1}. \end{align}
Whereas, for long dead times $\kappa\tau_\text{dead}\gg1$, then $\kappa t_{\text{int}}^*=W(-2/e^{2})+2\approx1.59$ is fixed and independent of $\bar N$, using the Lambert $W$ function. This corresponds to maximizing the ECQFI. The TOP SNR squared is then $2.6\,T\bar N/[\tau_\text{dead}\left(2 \bar n_{\text{th}}+1\right)]$. In either regime, the TOP SNR is unbounded in $\bar N$.

\begin{figure}
    \centering
    \includegraphics[width=\columnwidth]{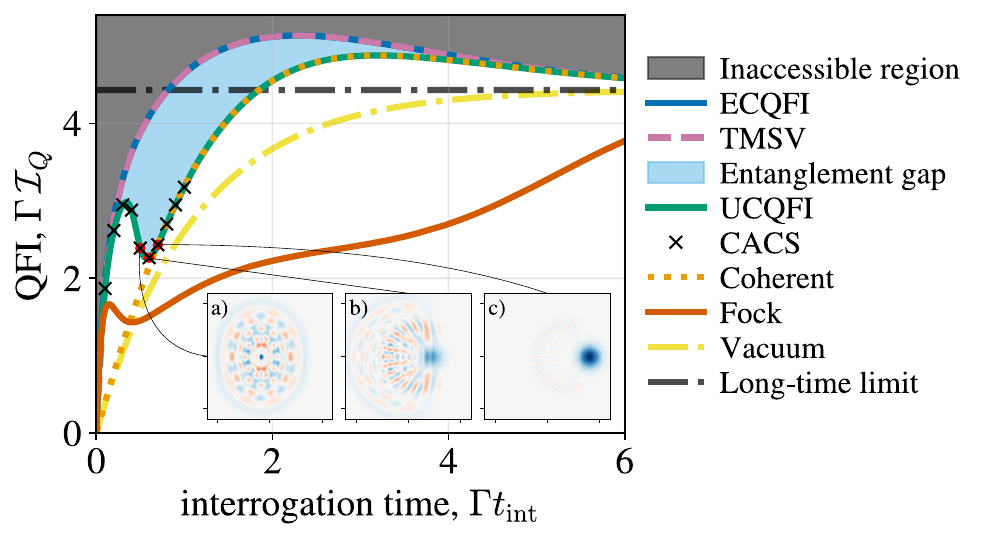}
    \caption{QFI of $\sqrt\gamma$ for fixed $\Gamma$ versus interrogation time $t_\text{int}$ for different initial states. The UCQFI line is an interpolated spline between the CACS results, the coherent QFI at later times, and Fock QFI at very short times~\cite{supplemental}. Results for $\bar N=8$ and $\Gamma/\gamma=20$. The three red dots among the CACS points indicate the states shown in the Wigner function plots for $\Gamma t_\text{int}$ equal to \textbf{a)} $0.5$, \textbf{b)}~$0.6$, and \textbf{c)} $0.7$.}     \label{fig:QFI vs time}
\end{figure} 
\vspace{2mm}
\textbf{Independent gain estimation.---} Let us now estimate the gain $\theta=\sqrt{\gamma}$ for a fixed loss $\Gamma$ in Eq.~\ref{eq:me}. These are independent gain and loss processes, where the gain $\gamma$ contributes solely $\mathcal{D}[\hat a^\dagger]$ and not also $\mathcal{D}[\hat a]$ to the master equation. This represents a three-level lasing process where the excited atomic state that the lasing transition goes to decays quickly to the ground state, or a blue-detuned optical pump driving an optomechanical system where the heating sideband decays quickly to vacuum. For zero dead time, this problem is equivalent to noise sensing if $\gamma\ll\kappa$ such that the contribution from $\gamma$ to $\Gamma$ can be ignored~\cite{gardner2025lindblad}. For finite dead time, however, this is surprisingly no longer true. Instead, since $\dot\eta\neq0$ and $\dot{\bar n}_\text{th}\neq0$, this problem combines noise sensing and loss estimation but also has some unique behavior. Since this scenario has been less studied previously, we explore it in more detail than the other cases.

The ECQFI in Eq.~\ref{eq:ECQFI} for this scenario is verbose and given in the End Matter, however, it simplifies in two relevant regimes. First, in the weak-signal limit of $\gamma\ll\Gamma$ for fixed $\bar N$ and $t_\text{int}$, the ECQFI is (cf.\ Eq.~\ref{eq:noise sensing}):
\begin{align}\label{eq:ECQFI_vanishing_gamma}        \IQ^\text{E} = \frac{4 (\bar N+1) \left(1-e^{-\Gamma  t_\text{int}}\right)}{\Gamma  \left[(\bar N+1) -\bar Ne^{-\Gamma  t_\text{int}}\right]}.
\end{align}
Second, in the high-energy limit of $\bar N\gg1$ for fixed $\gamma$ and $t_\text{int}$, the ECQFI is instead (cf.\ Eq.~\ref{eq:loss estimation}):
\begin{align}\label{eq:high energy limit}     \IQ^\text{E}=\frac{4\bar{N}\gamma t_\text{int}^{2}\left(\Gamma-\gamma\right)e^{-(\Gamma-\gamma)t_\text{int}}}{\left(\Gamma+\gamma\right)\left(1-e^{-(\Gamma-\gamma)t_\text{int}}\right)}.
    \end{align}
The order of limits here matters. Given the simultaneous conditions $\gamma\ll\Gamma$ and $\bar N(1-e^{-\Gamma t_\text{int}})\gg1$, the ECQFI becomes:
\begin{align}\label{eq:UB, simultaneous limit}
    \IQ^\text{E}    = \frac{4}{\Gamma}\left(1+\mathcal{Q}\right), \quad \mathcal{Q}:=\frac{\bar N \gamma \Gamma t_\text{int}^2}{e^{\Gamma t_\text{int}}-1}. 
\end{align}
Whether the weak signal limit (first term) or high energy limit (second term) dominates depends on the size of $\mathcal{Q}$. If $\mathcal{Q}\ll1$, then $\IQ^\text{E} = 4/\Gamma$ which corresponds to the $\bar N\to\infty$ limit of Eq.~\ref{eq:ECQFI_vanishing_gamma} and is bounded. Whereas, if $\mathcal{Q}\gg1$, then $\IQ^\text{E} = 4\mathcal{Q}/\Gamma$, which corresponds to the $\gamma\ll\Gamma$ limit of Eq.~\ref{eq:high energy limit} and grows linearly with $\bar N$, unlike noise sensing. Intuitively, fixing $\Gamma$ means that the signal has some characteristics of loss estimation. For short interrogation times $\Gamma t_\text{int}\ll1$, Eq.~\ref{eq:UB, simultaneous limit} assumes $\bar N \Gamma t_\text{int}\gg1$ such that $\mathcal{Q}=\bar N\gamma t_\text{int}$ then depends on the relative size of $\gamma/\Gamma\ll1$. If, nevertheless, $\mathcal{Q}\gg1$, then the TOP SNR squared is $(\text{SNR}_Q^{\text{E,TOP}})^2=\left(1-\sqrt{2\Gamma\tau_\text{dead}}\right)\times 4\bar N\gamma^2 T/\Gamma$ for $t^*_\text{int}=\sqrt{2\tau_\text{dead}/\Gamma}$, which is similar to loss estimation.

\begin{figure}
    \centering
    \includegraphics[width=\columnwidth]{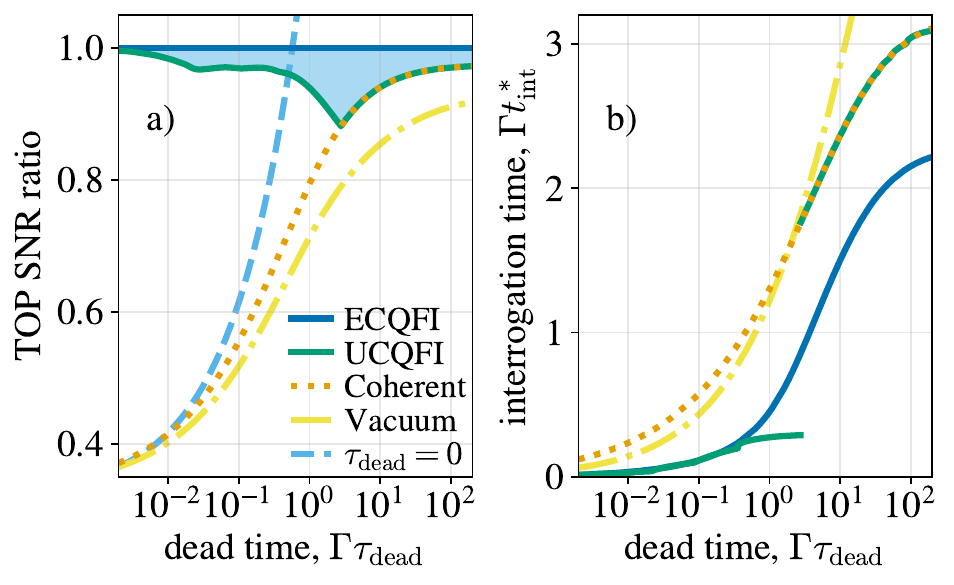}
    \caption{\textbf{a)}~Time-optimized (TOP) SNR relative to the TOP ECQFI SNR versus dead time for different initial states. We also show the entanglement gap and the vacuum protocol with no dead time. The roughness in the UCQFI curve comes from the interpolated spline, e.g.\ when transitioning between the Fock and CACS results at short dead times. \textbf{b)}~Optimal interrogation time $t_\text{int}^*$ versus dead time $\tau_\text{dead}$, showing the first-order discontinuity where the optimal unentangled state becomes the coherent state. Results for $\bar N=8$ and $\Gamma/\gamma=20$.} 
    \label{fig:dead_time} \end{figure}

The ECQFI is shown in Fig.~\ref{fig:QFI vs time} and TOP ECQFI SNR in Fig.~\ref{fig:dead_time}. By the high-energy limit in Eq.~\ref{eq:high energy limit} for long dead times $\Gamma \tau_{\text{dead}}\gg1$ and $\gamma\ll\Gamma$, the optimal interrogation time $t_{\text{int}}^*$ is the same as loss estimation below Eq.~\ref{eq:illumination, harsh approx} with $\kappa$ replaced by $\Gamma$. The maximum value of the ECQFI is $\IQ^\text{E}\approx2.59\bar{N}\gamma/\Gamma^{2}$ and the TOP SNR is $(\text{SNR}_Q^{\text{E,TOP}})^2\approx2.59\bar{N}\gamma^2 T/(\Gamma^{2}\tau_{\text{dead}})$. This is unlike noise sensing since it is unbounded in $\bar N$ regardless of $\tau_\text{dead}$.

We now study the entanglement gap between the ECQFI achieved with TMSV and the UCQFI achieved using unentangled states. 
We use a constrained alternating convex search (CACS)~\cite{gardner2025stochastic} to calculate the UCQFI numerically~\cite{supplemental}.
We show in Fig.~\ref{fig:QFI vs time} that an entanglement gap exists for finite $t_\text{int}$ and a given $\bar N$. We also observe a non-Gaussian to Gaussian transition in the optimal unentangled state with $t_\text{int}$. (This appears to be a property of the channel, not just this particular $\bar N$ value.) For short interrogation times ($\Gamma t_{\text{int}} \ll 1$), the ECQFI and UCQFI coincide with the Fock state QFI~\cite{gardner2025lindblad}. As $t_{\text{int}}$ increases, the optimal unentangled state is non-Gaussian but different from the Fock state. Then, the entanglement gap emerges, as the UCQFI is worse than the ECQFI and non-monotonic with $t_\text{int}$. Finally, the optimal unentangled state converges to the coherent state which is Gaussian. In the long-time limit, the QFI of all states converges to the thermal steady state QFI~\cite{supplemental}.
The non-monotonicity of the UCQFI with $t_\text{int}$ leads to a first-order discontinuity in the optimal interrogation time $t_\text{int}^*$ and a second-order discontinuity in the unentangled TOP SNR, as shown in Fig.~\ref{fig:dead_time}.
For long dead times, where the coherent state attains the UCQFI, its TOP SNR is $\sqrt{2.17\bar{N}\gamma^2 T/(\Gamma^{2}\tau_{\text{dead}})}$ for $t_{\text{int}}^*=\left[1+\sqrt{1+\Gamma/(\gamma\bar{N})}\right]/\Gamma$, given $\gamma \ll \Gamma$, $\bar N\gg1$, and $\gamma \bar N\gg\Gamma$~\cite{supplemental}. Comparing this to the TOP ECQFI SNR, the gap in the SNR is $9\%$ in the long dead time and high energy limits. Hence, the entanglement gap persists also in the long dead time limit, and it may relate to the gap discussed below Eq.~\ref{eq:coherent QFI, illumination}.

We observe in Fig.~\ref{fig:dead_time} that there is a threshold in the dead time around $\Gamma\tau_\text{dead}\sim1$ beyond which even the entangled scheme with dead time for a given energy is outperformed by the vacuum scheme without dead time. (The vacuum QFI is achieved using number-resolving measurements.)
The threshold and the gain in SNR over the vacuum below the threshold depend on the given energy. This result is relevant experimentally if the dead time for non-vacuum states is much longer than the time required to measure and reset the vacuum state.

\vspace{2mm}
\textbf{Applications.---} We now apply our results to solid-state mechanical resonators that can be prepared in nonclassical states of motion and used to sense forces, accelerations, and new physics~\cite{rahman_genuine_2025,Schrinski2023,omahen_ultracold_2026}. We focus on noise sensing as an example, relevant for ultralight dark matter or high-frequency gravitational wave searches.
 Following Ref.~\cite{rahman_genuine_2025}, we consider $\bar N\approx 5$, $1/\kappa \approx 100\,\unit{\micro\second}$, and $\tau_{\rm dead}\approx 200\,\unit{\micro\second}$.
Let us assume that $1/\gamma=100\,\unit{\milli\second}$.
Using the ECQFI in Eq.~\ref{eq:noise sensing}, the TOP SNR is $\sqrt{T/78\,\unit{\milli\second}}$ for $t^*_\text{int}=59\,\unit{\micro\second}$. This is 20\% larger than the TOP SNR for the vacuum state attained at $t^*_\text{int}=150\,\unit{\micro\second}$ for the same dead time, and the optimal ECQFI per cycle is 83\% of the high energy limit. If active resets could halve the dead time, then the TOP SNR increases to $\sqrt{T/47\,\unit{\milli\second}}$ for $t^*_\text{int}=41\,\unit{\micro\second}$. A similar analysis can be done for the other parameterizations.

\vspace{2mm}
\textbf{Conclusions.---} We have studied the metrological impact of finite dead time arising from state preparation, measurement, and reset operations. For prepare-wait-measure-reset metrology for stochastic signal sensing, we have shown how the optimal interrogation time to maximize the SNR depends on the dead time. We proved that preparing TMSV is optimal for sensing any parameter of the thermal loss channel, even at finite energy. This includes noise sensing, quantum illumination, and sensing a gain independent of a loss. We showed that whether the SNR is bounded and an entanglement gap exists can change dramatically depending on the parameterization. Experimental design should account for the optimal strategy changing with the dead time. We applied these results to bulk acoustic wave resonators.

We assumed for simplicity that the dead time is independent of the chosen initial state and measurement. In practice, however, the dead time depends on the chosen strategy, e.g.\ whether entanglement or high non-Gaussianity are used, and thus determining the TOP strategy involves a further tradeoff. Experimentally-informed future work should determine how the dead time depends on the chosen strategy and calculate the resulting TOP SNR. Future work should also examine the general behavior of the entanglement gap, loss estimation with finite interrogation time, multi-parameter channel estimation, stochastic signals with finite coherence time $\tau_\phi$~\cite{shi2025quantum,freiman2025quantum} or finite samples, and protocols with control~\cite{shin2026heisenberg}.

\vspace{2mm}
\textbf{Data availability.---} The data that support the findings of this article are openly available~\cite{repo}.

\vspace{2mm}
\textbf{Acknowledgments.---} We thank Aashish Clerk, Roberto Di Candia, Nadav Katz, Harel Radia, and Cindy Regal for their helpful advice. 
J.W.G.\ acknowledges support from the ARO (W911NF-23-1-0077), ARO MURI (W911NF-21-1-0325), AFOSR MURI (FA9550-21-1-0209, FA9550-23-1-0338), DARPA (HR0011-24-9-0359, HR0011-24-9-0361), NSF (ERC-1941583, OMA-2137642, OSI-2326767, CCF-2312755, OSI-2426975), Packard Foundation (2020-71479), and DOE (DE-AC02-06CH11357). T.G.\ acknowledges funding provided by the Quantum Science and Technology early-career fellowship of the Israel Council for Higher Education and ISF Grant No.\ 3302/25.
M.F. was supported by the Swiss National Science Foundation Ambizione Grant No. 208886, and by The Branco Weiss Fellowship -- Society in Science, administered by the ETH Z\"{u}rich.

\begin{widetext}

\setlength{\tabcolsep}{6pt}
\begin{table*}[ht!]
\begin{tabular}{@{}llllllll@{}}
\toprule
Signal scenario & Parameter, $\theta$ & Fixed & Short dead time & $t_\text{int}^*$ & Long dead time & $t_\text{int}^*$ & Gap \\ \midrule
Coherent signal & $\Omega_0$ & $\kappa$ & $[1-(3\kappa\tau_\text{dead})^{2/3}/4]\times\frac{4\Omega_0^2T}{\kappa}$ & $2\left(\frac{3\tau_\text{dead}}{\kappa^2}\right)^{1/3}$ & $\frac{16\Omega_0^2T}{\kappa^2 \tau_\text{dead}}$ & $\frac{2}{\kappa}\log(\kappa \tau_\text{dead})$ & $\times$  \\[2pt]
Noise sensing & $\sqrt{\gamma}$ & $\kappa$ & $\left(1-2\sqrt{\kappa\bar N\tau_\text{dead}}\right)\times4 \gamma T (\bar N+1)$ & $\sqrt{\frac{\tau_\text{dead}}{\kappa\bar N}}$ & $\frac{4\gamma T}{\kappa \tau_\text{dead}}$ & $\sqrt{\frac{\tau_\text{dead}}{\kappa\bar N}}$ & $\times$  \\[2pt]
Loss estimation & $\sqrt{\kappa}$ & $\bar n_\text{th}$ & $\left(1-\sqrt{2\kappa\tau_\text{dead}}\right)\times\frac{4\kappa T\bar N}{2 \bar n_{\text{th}}+1}$ & $\sqrt{\frac{2\tau_\text{dead}}{\kappa}}$ & $\frac{2.6\,T\bar N}{\tau_\text{dead}\left(2 \bar n_{\text{th}}+1\right)}$ & $\frac{1.59}{\kappa}$ & $\checkmark$  \\[2pt]
Independent gain & $\sqrt{\gamma}$ & $\Gamma$ & $\left(1-\sqrt{2\Gamma\tau_\text{dead}}\right)\times\frac{4\bar N\gamma^2 T}{\Gamma}$ & $\sqrt{\frac{2\tau_\text{dead}}{\Gamma}}$ & $\frac{2.6\bar{N}\gamma^2 T}{\Gamma^{2}\tau_{\text{dead}}}$ & $\frac{1.59}{\Gamma}$ & $\checkmark$ \\ \bottomrule
\end{tabular}\caption{TOP SNR squared and optimal interrogation time $t_\text{int}^*$ for short and long dead times for different signal scenarios with different quantities fixed. The precise definition of what short and long dead time means in each context differs slightly and is provided in the main text. We also indicate whether an entanglement gap is observed at high energies, at least for some parameters.} \label{tab:comparison}
\end{table*}

\end{widetext}

\begingroup
\renewcommand{\addcontentsline}[3]{}\section*{End Matter}
\endgroup
We now discuss the ECQFI in some more technical detail. For the thermal loss channel, the ECQFI in Eq.~\ref{eq:ECQFI} is: \begin{align}
    \IQ^\text{E}=\frac{c_1\bar N^2+c_2\bar N+c_3}{c_4\bar N+c_5},
\end{align}
where the verbose coefficients are:
\begin{align}\label{eq:coeffs, EM}
    c_1 &= (1-\eta ) \bar n_{\text{th}} (\bar n_{\text{th}}+1)\dot{\eta }^2,  \\
    c_2 &= \bar n_{\text{th}} (\bar n_{\text{th}}+1) (2 \eta  \bar n_{\text{th}}+1)\dot{\eta }^2 
        \nonumber\\&\quad- 4 (1-\eta) \eta  \bar n_{\text{th}} (\bar n_{\text{th}}+1) \dot{\eta } \dot{\bar n}_{\text{th}}
        \nonumber\\&\quad+(1-\eta)^2 \eta  (2 \bar n_{\text{th}}+1) \dot{\bar n}_{\text{th}}^2, \nonumber\\
    c_3 &= \eta  (\bar n_{\text{th}}+1) [\bar n_{\text{th}}\dot{\eta } -(1-\eta) \dot{\bar n}_{\text{th}}]^2, \nonumber\\
    c_4 &= (1-\eta)^2 \eta \bar n_{\text{th}} (\bar n_{\text{th}}+1) (2 \bar n_{\text{th}}+1), \nonumber\\
    c_5 &= (1-\eta) \eta \bar n_{\text{th}} (\bar n_{\text{th}}+1) [(1-\eta) \bar n_{\text{th}}+1],\nonumber
\end{align}
and we define $\dot\eta=\partial_\theta\eta$ and $\dot{\bar n}_\text{th}=\partial_\theta \bar n_\text{th}$ where $\partial_\theta=\frac{\partial}{\partial\theta}$. Here, the coefficients $c_4$ and $c_5$ in the denominator do not depend on the choice of parameter $\theta$ to estimate. In the numerator, $c_1\propto\dot\eta^2$ is only nonzero if there is a loss estimation component to the parameter. When this is the case, then the ECQFI can increase linearly in $\bar N$ indefinitely, leading to unbounded TOP SNR even for finite dead time. The other coefficients $c_2$ and $c_3$ in the numerator depend on both $\dot\eta$ and $\dot{\bar n}_\text{th}$. If the parameter has only a noise sensing component $\dot{\bar n}_\text{th}\neq0$ with $\dot\eta=0$, then $c_2\propto \dot{\bar n}_\text{th}^2$ and $c_3\propto \dot{\bar n}_\text{th}^2$ and although the ECQFI can increase linearly with $\bar N$ initially, it eventually saturates to some finite high-energy limit.
This QFI result holds for $0<\eta<1$, where the $\eta=0$ and $\eta=1$ cases are understood in the limiting sense. We also assume that $\bar n_\text{th}>0$ if $\dot{\bar n}_\text{th}\neq0$ but otherwise allow $\bar n_\text{th}\geq0$ if $\dot{\bar n}_\text{th}=0$.

In the long interrogation time limit of $t_\text{int}\to\infty$, the ECQFI can be verified using an independent method~\cite{supplemental}. The result is then the sum of the loss estimation QFI from Ref.~\cite{PhysRevLett.118.070803} times $\dot\eta^2/(4\eta)$ and the QFI of the thermal steady state with respect to $\theta$ (which is proportional to $\dot{\bar n}_\text{th}^2$). Thus, an entanglement gap exists if and only if $\dot\eta^2/(4\eta)\neq0$ in this limit.

The ECQFI in Eq.~\ref{eq:ECQFI} for independent gain estimation of $\theta=\sqrt\gamma$ for fixed $\Gamma$ is also verbose as the coefficients are: 
\begin{align}\label{eq:Scenario 1 coefficients}
 c_1 &= 4 \gamma  t^2 (\Gamma - \gamma)^3 \eta \left(1-\eta\right), \\
 c_2 &= 4 \Gamma (\Gamma + \gamma)
 \nonumber\\&+4 \eta^2 \left[\Gamma  (\Gamma + \gamma)+2 \gamma ^2 t^2 (\Gamma - \gamma)^2+4 \gamma  \Gamma  t (\Gamma -\gamma )\right]
 \nonumber\\&-4 \eta \left[2 \Gamma  (\Gamma + \gamma)-\gamma  t^2 (\Gamma - \gamma)^3+4 \gamma  \Gamma  t (\Gamma -\gamma )\right] ,\nonumber\\
 c_3 &= 4 \left[\Gamma - \eta (\gamma  t (\Gamma - \gamma)+\Gamma )\right]^2 ,\nonumber\\
 c_4 &= (\Gamma - \gamma)^2 (\Gamma + \gamma) \left(1-\eta\right)^2 ,\nonumber\\
 c_5 &= (\Gamma -\gamma)^2 \left(1-\eta\right) \left(\Gamma - \gamma  \eta\right) \nonumber
 ,
\end{align}
which are not the same as those in Eq.~\ref{eq:coeffs, EM} as we have canceled some common factors in the numerator and denominator. Nevertheless, they give the same ECQFI. This yields Eq.~\ref{eq:ECQFI_vanishing_gamma} as $(c_2\bar N+c_3)/(c_4\bar N+c_5)$ for $\gamma\ll\Gamma$ and Eq.~\ref{eq:high energy limit} as $c_1\bar N/c_4$ for $\bar N\gg1$. These two limits are naive as they only keep the leading order term in either case. The simultaneous limit in Eq.~\ref{eq:UB, simultaneous limit} is $(c_1\bar N+c_2)/c_4$ provided that $\gamma\ll\Gamma$ and $\bar N(1-e^{-\Gamma t_\text{int}})\gg1$ to drop $c_3$ and $c_5$, as these conditions guarantee that $c_2\bar N/c_3\gg1$ and $c_4\bar N/c_5\gg1$. We then expand $c_1$, $c_2$, and $c_4$ to leading order in $\gamma\ll\Gamma$ to reach Eq.~\ref{eq:UB, simultaneous limit}. This demonstrates that there is a transition between the weak signal and high energy regimes depending on $\mathcal{Q}$.

For ease of reference, we summarize the results in the main text about the scaling of the optimal interrogation time $t_\text{int}^*$ and the TOP SNR with the dead time $\tau_\text{dead}$ in Table~\ref{tab:comparison} for the various cases studied. The precise conditions on each of the results, given in the main text, are omitted for simplicity. However, these should be noted because the order of limits matters.

\clearpage
\newpage

\widetext
    
\setcounter{page}{1}
\setcounter{figure}{0}
\setcounter{table}{0}
\setcounter{section}{0}
\setcounter{equation}{0}

\renewcommand{\thetable}{S\arabic{table}}  
\renewcommand{\thepage}{\arabic{page}}  
\renewcommand{\thefigure}{S\arabic{figure}}
\renewcommand{\theHfigure}{S\arabic{figure}}
\renewcommand{\theequation}{S\arabic{equation}}

\setcounter{secnumdepth}{3} 
\renewcommand{\thepage}{\arabic{page}}
\renewcommand{\thefigure}{S\arabic{figure}}
\renewcommand{\thetable}{S\arabic{table}}
\renewcommand{\theequation}{S\arabic{equation}}
\renewcommand{\thesection}{S\arabic{section}}
\renewcommand{\thesubsection}{S\arabic{section}.\arabic{subsection}}
\renewcommand{\thesubsubsection}{S\arabic{section}.\arabic{subsection}.\arabic{subsubsection}}

\renewcommand{\theHfigure}{S\arabic{figure}}
\renewcommand{\theHtable}{S\arabic{table}}
\renewcommand{\theHequation}{S\arabic{equation}}
\renewcommand{\theHsection}{S\arabic{section}}

\makeatletter
\renewcommand{\p@section}{}
\renewcommand{\p@subsection}{}
\renewcommand{\p@subsubsection}{}
\makeatother

{\centering\textbf{Supplemental Material for:} \\} 
{\centering\textbf{Stochastic signal sensing with finite dead time at the fundamental quantum limit}\\} 
\normalsize
\vspace{.3cm}
{\centering James W. Gardner\,$^{1,}$$^\ast$, Tuvia Gefen\,$^{2,}$$^\dagger$, and Matteo Fadel\,$^{3,}$$^\ddagger$\\
\vspace{2mm}
\textit{$^1$ Chicago Quantum Institute, Pritzker School of Molecular Engineering, University of Chicago, Illinois 60637, USA\\}
\textit{$^2$ Racah Institute of Physics, The Hebrew University of Jerusalem, Jerusalem 91904, Givat Ram, Israel\\}
\textit{$^3$ Department of Physics, ETH Z\"{u}rich, 8093 Z\"{u}rich, Switzerland\\}
\vspace{2mm}
{\footnotesize $^\ast$ \color{blue} jamesgardner@uchicago.edu\\}
{\footnotesize $^\dagger$ \color{blue} tuvia.gefen@mail.huji.ac.il\\}
{\footnotesize $^\ddagger$ \color{blue} fadelm@phys.ethz.ch\\}
}

\suppressfloats

\makeatletter
\begingroup
\let\SM@orig@contentsline\contentsline
\def\contentsline#1#2#3#4{  \def\SM@tocentry{#2}  \def\SM@endmatterentry{\numberline {}End Matter}  \ifx\SM@tocentry\SM@endmatterentry
  \else
    \SM@orig@contentsline{#1}{#2}{#3}{#4}  \fi
}\tableofcontents
\endgroup
\makeatother

\clearpage
\newpage
\section{List of symbols and abbreviations used in the main manuscript}

\begin{table*}[h!]
\centering
\makeatletter
\let\@mkpream\arxiv@mkpream
\makeatother
\small
\setlength{\tabcolsep}{4pt}
\renewcommand{\arraystretch}{1.15}
\begin{tabular}{@{}p{0.1\textwidth}p{0.38\textwidth}|p{0.1\textwidth}p{0.38\textwidth}@{}}
\toprule
Symbol & Meaning & Symbol & Meaning \\
\midrule
$\hat a,\hat a^\dagger$ & bosonic annihilation and creation operators & $\mathcal{C}_{\eta,\bar n_\text{th}}$ & Gaussian thermal loss channel \\
$\hat n=\hat a^\dagger\hat a$ & number operator & $\vec\mu$ & Gaussian-state mean vector \\
$\hat\rho$ & density operator & $\Sigma$ & Gaussian-state covariance matrix \\
$\hat\rho_0$ & initial density operator & $\Sigma_\text{th}$ & thermal covariance matrix \\
$\hat\rho(t_\text{int})$ & state after interrogation time & $I_2$ & $2\times2$ identity matrix \\
$\hat H$ & Hamiltonian & $\theta$ & generic parameter to be estimated \\
$\hbar$ & reduced Planck constant & $\Delta\theta$ & standard deviation of an unbiased estimator of $\theta$ \\
$\mathcal{D}[\hat A]\hat\rho$ & Lindblad dissipator with jump operator $\hat A$ & $\text{SNR}$ & signal-to-noise ratio \\
$\Omega(t)$ & complex-valued drive or stochastic signal & $\text{SNR}_C$ & CFI Cram\'er--Rao SNR bound \\
$\Omega_0$ & coherent signal amplitude & $\text{SNR}_Q$ & QFI Cram\'er--Rao SNR bound \\
$\phi$ & signal phase & $\text{SNR}_Q^{j,\text{TOP}}$ & time-optimized QFI SNR bound for strategy $j$ \\
$\omega$ & angular frequency & $\IC$ & classical Fisher information (CFI) \\
$S_\Omega(\omega)$ & power spectral density of $\Omega(t)$ & $\IQ$ & quantum Fisher information (QFI) \\
$\tau_\phi$ & signal correlation/coherence time & $\IQ^\text{E}$ & entanglement-assisted channel QFI (ECQFI) \\
$\kappa$ & environmental damping or loss rate & $\IQ^\text{U}$ & unentangled channel QFI (UCQFI) \\
$\gamma$ & gain rate & $j\in\{\text{E},\text{U}\}$ & strategy label: entanglement-assisted or unentangled \\
$\Gamma$ & effective loss rate & $\hat L$ & symmetric logarithmic derivative \\
$\eta=e^{-\kappa t_\text{int}}$ & channel transmissivity & $p_j$ & probability of measurement outcome $j$ \\
$\bar n_\text{th}$ & thermal bath occupation number & $\{\hat E_j\}_j$ & measurement effects \\
$t$ & dynamical time coordinate & $\partial_\theta$ & derivative with respect to $\theta$ \\
$t_\text{int}$ & interrogation time & $\dot\eta,\dot{\bar n}_\text{th}$ & derivatives $\partial_\theta\eta$ and $\partial_\theta\bar n_\text{th}$ \\
$t_\text{int}^*$ & optimal interrogation time & $\bar N$ & average particle-number constraint per mode \\
$\tau_\text{dead}$ & dead time for preparation, measurement, and reset & $r$ & squeezing parameter \\
$T$ & total experiment time & $\ket{\psi}$ & single-mode pure probe state \\
$M$ & number of experimental repetitions & $\ket{\Psi}$ & two-mode pure probe state including an ancilla \\
$\mathcal{E}_{t_\text{int},\phi}$ & coherent-signal Gaussian channel & $c_n$ & coefficients appearing in the ECQFI expression \\
$\bar{\mathcal{E}}_{t_\text{int}}$ & phase-averaged coherent-signal channel & $\mathcal{Q}$ & parameter for the independent-gain crossover \\
$\mathcal{L}_\eta$ & vacuum loss channel with transmissivity $\eta$ & $W$ & Lambert $W$ function \\
$\mathcal{T}_\alpha$ & displacement channel with displacement $\alpha$ & TMSV & two-mode squeezed vacuum state \\
$\alpha$ & complex displacement amplitude & CACS & constrained alternating convex search \\
 &  & TOP & time-optimized \\
\bottomrule
\end{tabular}
\label{tab:symbols-main}
\end{table*}

\clearpage
\newpage
\section{Models of a stochastic signal in the presence of loss}

We now prove that the different models for the stochastic signal $\Omega(t)$ in the presence of loss in Eq.~\ref{eq:Hamiltonian} in the main text reduce, in appropriate limits, to the thermal loss channel $\mathcal{C}_{\eta,\bar n_\text{th}}$ for a given interrogation time. For certain parameter regimes, these models are furthermore described by the master equation dynamics in Eq.~\ref{eq:me} for all times.

Our approach is to first show that we can focus on the random displacement channel given by the signal $\Omega(t)$ and the memory kernel given the loss rate $\kappa$. In particular, it suffices to show that this random displacement channel is approximately the Gaussian isotropic additive noise channel. We then show this for the different models for the stochastic signal given in the main text.

\subsection{Random displacement channels}
Let us start by solving the master equation evolution in Eq.~\ref{eq:Hamiltonian} for a particular classical trajectory $\Omega(t)$ with probability process functional $\pi[\Omega(t)]$. This can be freely interpreted as our Bayesian prior knowledge about some deterministic signal $\Omega$ or as the frequentist likelihood of a stochastic source to yield $\Omega$. We can imagine discretising the time domain into short time bins to make this into a regular multivariate probability distribution if we want to avoid functionals and the signal is band-limited.

The Heisenberg-picture time evolution in Eq.~\ref{eq:Hamiltonian} for some operator $\hat O(t)$ for a given $\Omega(t)$ is:
\begin{align*}
    \deriv{\hat O(t)}{t}
    &=+\frac{i}{\hbar}[\hat H,\hat O(t)]+\kappa \left(\hat a(t)^\dagger\hat O(t)\hat a(t)-\frac{1}{2}\{\hat a(t)^\dagger \hat a(t), \hat O(t)\}\right), \qquad
    \hat H=\hbar \left(\Omega(t) \hat a^\dagger + \Omega^\ast(t) \hat a \right),
\end{align*}
such that the evolution for $\hat a(t)$ itself is:
\begin{align*}
    \deriv{\hat a(t)}{t}
    &=+\frac{i}{\hbar}[\hat H,\hat a(t)]+\kappa \left(\hat a(t)^\dagger\hat a(t)\hat a(t)-\frac{1}{2}\{\hat a(t)^\dagger \hat a(t), \hat a(t)\}\right)
                    =-i\Omega(t)-\frac{\kappa}{2}\hat a(t),
\end{align*}
the solution of which given the initial condition $\hat a(0)$ is:
\begin{align}\label{eq:a solution to a}
    \hat a(t)=\hat a(0)e^{-\kappa t/2}-i\int_{0}^{t}\text{d}s\;e^{-\kappa\left(t-s\right)/2}\Omega(s)=\sqrt{\eta}\;\hat a(0)+\alpha_{t},\qquad
    \alpha_{t}:=-i\int_{0}^{t}\text{d}s\;e^{-\kappa\left(t-s\right)/2}\Omega(s),
\end{align}
where recall that $\eta=e^{-\kappa t}$ is the power transmissivity.
Here, $\alpha_{t}$ is a function of time $t$, the loss $\kappa$, and the trajectory $\Omega(s)$.
If the signal $\Omega(t)=\Omega_0e^{i\phi}$ is time-independent, as in the coherent, phase-locked, and tuned signal case, then $\alpha_{t}=-i\frac{2\Omega_0}{\kappa}(1-\sqrt{\eta})e^{i\phi}$.
As another deterministic example, if $\Omega(t)=\Omega _0 e^{i (\omega t + \phi )}$ where $\omega$ is the sideband/detuning frequency since we are in the rotating frame, then $\alpha_{t}
=-i\frac{2 \Omega _0}{\kappa + i 2\omega }\left(e^{i \omega t} - \sqrt{\eta}\right) e^{i \phi}$ so if we are far detuned $\abs{\omega}\gg\kappa/2$ then the signal is approximately $|{\alpha_{t}}|\approx \frac{\Omega _0}{\abs\omega }|e^{i \omega t} - \sqrt{\eta}|\leq\frac{\Omega _0}{\abs\omega }(1+\sqrt{\eta})$ which is small if furthermore $\abs{\omega}\gg\Omega_0$. Note that this result about the signal being small does not depend on the relative size of $\kappa/2$ and $\Omega_0$.

For a given time $t$ and trajectory $\Omega$, this evolution is equivalent to the channel $\mathcal{A}_{t,\Omega}:=\mathcal{T}_{\alpha_{t}}\circ\mathcal{L}_{\eta}$ corresponding to a vacuum loss followed by a unitary displacement by $\alpha_{t}$. (We label this general channel as $\mathcal{A}_{t,\Omega}$ to avoid confusion with $\mathcal{E}_{t_\text{int},\phi}$, which is its specification to the time-independent coherent, phase-locked signal case above.) We now want to marginalize over the probability density $\pi[\Omega]$ of the particular trajectory $\Omega$ occurring. Note that we marginalize at a particular time $t$ when we make a measurement and want to describe the results of the measurement with respect to the classical average over the ensemble of trajectories. We distinguish this from quantum projection noise and from independently marginalizing the signal at intermediate times which represents a different physical system (our jointly marginalized channel is not necessarily CP-divisible). Marginalizing at a particular time $t$ leads to the following channel:
\begin{align}\label{eq:marginalized sol}
    \bar{\mathcal{A}}_{t,\pi}:=\int\text{D}\pi[\Omega]\;\mathcal{A}_{t,\Omega}=\int\text{D}\pi[\Omega]\;\mathcal{T}_{\alpha_{t}}\circ\mathcal{L}_{\eta}
    =\bar{\mathcal{T}}_{\pi_t}\circ\mathcal{L}_{\eta},\qquad
    \bar{\mathcal{T}}_{\pi_t}:=\int\text{D}\pi[\Omega]\;\mathcal{T}_{\alpha_{t}}=\int\text{d}^2\alpha_{t}\;\pi_t(\alpha_{t})\;\mathcal{T}_{\alpha_{t}},
\end{align}
where we integrate the functional measure $\text{D}\pi[\Omega]$ times the channel over all possible trajectories. We change variables to the 2D phase-space probability distribution $\pi_t(\alpha_{t})$ of the real/imaginary parts of the complex displacement $\alpha_{t}$ for a given time $t$ and trajectory $\Omega$. The push-forward measure becomes $\text{d}^2\alpha_{t}\;\pi(\alpha_{t})$ as the conversion is:
\begin{align}\label{eq:pushforward measure}
    \pi_t(\beta) = \int\text{D}\pi[\Omega]\;\delta(\beta-\alpha_{t}), \qquad \text{such that}\qquad \int\text{d}^2\beta\;\pi_t(\beta) f(\beta) = \int\text{D}\pi[\Omega]\;f(\alpha_{t}),\quad \forall f:\mathbb{C}\to\mathbb{C}.
\end{align}
This mapping is non-unique for a particular time $t$ but is unique on some time interval provided that $\Omega$ is continuous.

Let us define the generic single-mode random displacement channel $\bar{\mathcal{T}}_{p}$ for a given 2D phase-space distribution $p:\mathbb{C}\to\mathbb{R}_{\geq0}$:
\begin{align}\label{eq:random displacement}
    \bar{\mathcal{T}}_{p}[\hat\rho]
    =\int\text{d}^2\alpha\;p(\alpha)\mathcal{T}_{\alpha}[\hat\rho]
        =\int\text{d}^2\alpha\;p(\alpha)\hat D(\alpha)\hat\rho \hat D^\dagger(\alpha).
\end{align}
This probability distribution $p$ leads to classical expectation values $\evTextSub{p}{f(\alpha)}:=\int\text{d}^2\alpha\;p(\alpha)f(\alpha)$ and variances:
\begin{align*}
    \varSub{p}{f(\alpha)}:=\evTextSub{p}{\abs{f(\alpha)-\evTextSub{p}{f(\alpha)}}^2}=\evTextSub{p}{\abs{f(\alpha)}^2}-\abs{\evTextSub{p}{f(\alpha)}}^2,
\end{align*}
for any complex-valued test function $f:\mathbb{C}\to\mathbb{C}$. Note that $\evTextSub{p}{f(\alpha)}\in\mathbb{C}$ but $\varSub{p}{f(\alpha)}\geq0$.
The random displacement channel $\bar{\mathcal{T}}_{p}$ is a Gaussian channel if and only if $p$ is a 2D Gaussian distribution. For example, $p$ Gaussian with zero mean and covariance matrix $\Sigma=\text{diag}(\sigma_x^2,\sigma_p^2+\sigma^2)$ gives the stochastic signal model with asymmetric classical noise studied for noise sensing in Ref.~\cite{gardner2025stochastic}. We give a non-Gaussian example later below.

Returning to our scenario, in Eq.~\ref{eq:marginalized sol}, the channel $\bar{\mathcal{T}}_{\pi_t}$ is the single-mode random displacement channel for $p=\pi_t$. The dependence of $\alpha_{t}$ on $t$ and $\Omega$ in Eq.~\ref{eq:a solution to a} is baked into $\pi_t$ in Eq.~\ref{eq:pushforward measure}. 
If $\Omega(t)$ is a Gaussian complex-valued random process, i.e.\ $\pi[\Omega]$ is a Gaussian functional, then $\alpha_{t}$ is also Gaussian since it is a linear combination of Gaussian random variables in Eq.~\ref{eq:a solution to a}, i.e.\ $\pi_t(\alpha)$ is a 2D Gaussian distribution in Eq.~\ref{eq:pushforward measure}. Meanwhile, the deterministic (unitary displacement) examples above are singular instances of the random displacement channel where $\pi_t(\alpha)=\delta(\alpha-\beta)$ for some $\beta$.

In the following subsections, we will show that certain distributions $\pi[\Omega]$ of random processes $\Omega(t)$ lead to the Gaussian thermal loss channel $\mathcal{C}_{\eta,\bar n_\text{th}}$ under suitable approximations. The key is simply to find the resulting distribution $\pi_t$ of random displacements and show that it is approximately the isotropic 2D Gaussian distribution.
This is because of the following decomposition of the thermal loss channel, $\mathcal{C}_{\eta,\bar n_\text{th}}=\mathcal{N}_\nu\circ\mathcal{L}_\eta$, into a loss channel $\mathcal{L}_\eta$ followed by the isotropic Gaussian random displacement channel $\mathcal{N}_\nu$ where $\nu:=(1-\eta)\bar n_\text{th}$~\cite{holevo2007one}. To give more details, $\mathcal{N}_\nu$ is the Gaussian isotropic additive noise channel which acts as $\Sigma\to\Sigma+\nu\,\text{diag}(1,1)$ and leaves $\vec\mu$ invariant.
Thus, $\mathcal{N}_\nu=\bar{\mathcal{T}}_{\varpi}$ is a random displacement channel with isotropic Gaussian kernel $\varpi$ such that $\evTextSub{\varpi}{\alpha}=0$ and $\varSub{\varpi}{\alpha}=\text{E}_{\varpi}[\abs{\alpha}^2]=\nu$, which is circular since $\text{E}_{\varpi}[\alpha^2]=0$. We use a new notation for this particular random displacement channel since it is so common and is also important to our proof.
Therefore, we have that $\bar{\mathcal{A}}_{t,\pi}=\bar{\mathcal{T}}_{\pi_t}\circ\mathcal{L}_{\eta}$ from Eq.~\ref{eq:marginalized sol} and now $\mathcal{C}_{\eta,\bar n_\text{th}}=\mathcal{N}_\nu\circ\mathcal{L}_\eta$.
These decompositions imply that, whenever $\bar{\mathcal{T}}_{\pi_t}\approx \mathcal{N}_\nu$ (i.e.\ $\pi_t\approx\varpi$) for some $\nu$, then our channel can be approximated as a thermal loss channel $\bar{\mathcal{A}}_{t,\pi}\approx\mathcal{C}_{\eta,\bar n_\text{th}}$.
Note that the vacuum loss channel $\mathcal{L}_\eta$ can be left exact in this comparison, e.g.\ there is no need for us to assume that $\kappa t\ll1$. The loss rate $\kappa$ still appears, however, in the definition of $\alpha_{t}$ in Eq.~\ref{eq:a solution to a} and thus in how we convert $\pi[\Omega]$ to $\pi_t(\alpha)$ in Eq.~\ref{eq:pushforward measure}. 

We now consider the different stochastic signal models in the main text and show that $\bar{\mathcal{T}}_{\pi_t}\approx \mathcal{N}_\nu$ to complete the proof.

\subsection{Coherent signal, not phase locked with the source}
The first stochastic signal model in the main text is the coherent, not phase locked signal $\Omega(t)=\Omega_0e^{i\phi}$ with $\phi$ unknown shot-to-shot. This channel is $\bar{\mathcal{E}}_{t_\text{int}}=\frac{1}{2\pi}\int_0^{2\pi}\text{d}\phi\;\mathcal{E}_{t_\text{int},\phi}$ where $\mathcal{E}_{t_\text{int},\phi}=\mathcal{T}_{\alpha}\circ\mathcal{L}_{\eta}$. In our general notation introduced above, $\bar{\mathcal{A}}_{t,\pi}=\bar{\mathcal{E}}_{t_\text{int}}$ with $\bar{\mathcal{T}}_{\pi_t}=\frac{1}{2\pi}\int_0^{2\pi}\text{d}\phi\;\mathcal{T}_{\alpha}$ for $\alpha=-i\frac{2\Omega_0}{\kappa}(1-\sqrt{\eta})e^{i\phi}$ which is a function of time through $\eta=e^{-\kappa t}$. This is consistent with the 2D random displacement channel $\bar{\mathcal{T}}_{\pi_t}$ for $\pi_t(\beta)=\frac{1}{2\pi\abs{\alpha}}\delta(\abs{\beta}-\abs{\alpha})$ which is non-Gaussian in general.

We now take the weak signal limit such that the non-Gaussianity of $\bar{\mathcal{T}}_{\pi_t}$ is negligible. This fixed-radius random displacement channel using Eq.~\ref{eq:random displacement} is:
\begin{equation}\label{eq:marginal phase displacement}
\bar{\mathcal{T}}_{\pi_t}[\hat \rho]
=\frac{1}{2\pi}\int_{0}^{2\pi}\text{d}\phi\; \hat D(\alpha)\hat \rho \hat D^\dagger(\alpha), \qquad \alpha=-i\frac{2\Omega_0}{\kappa}(1-\sqrt{\eta})e^{i\phi} \;.
\end{equation}
Here, the displacement is $\hat D(\alpha)=\exp(\alpha \hat a^\dagger-\alpha^* \hat a)=\exp(\hat X)$ with $\hat X=\alpha \hat a^\dagger-\alpha^* \hat a$ such that we can then expand:
\begin{align*}
\hat D(\alpha)\hat \rho \hat D^\dagger(\alpha)
= \hat \rho + [\hat X,\hat \rho] + \frac{1}{2}[\hat X,[\hat X,\hat \rho]] + \mathcal{O}(|\alpha|^3),
\end{align*}
where the expansion is in approximately $\abs{\alpha}\sqrt{\bar N+1}=\frac{2\Omega_0}{\kappa}(1-\sqrt{\eta})\sqrt{\bar N+1}\ll1$ being small since $\hat X^2$ contains terms like $\hat a\hat a^\dagger$.
Averaging over \(\phi\) uniformly in Eq.~\ref{eq:marginal phase displacement} then yields that all terms with a net complex phase vanish:
\[
\langle \alpha \rangle_\phi = 0, \qquad 
\langle \alpha^2 \rangle_\phi = 0, \qquad 
\langle (\alpha^*)^2 \rangle_\phi =0, \qquad \text{but}\qquad
\langle |\alpha|^2 \rangle_\phi = |\alpha|^2.
\]
Hence the first-order term \(\langle[\hat X,\hat \rho]\rangle_\phi\) vanishes, and only the second-order term contributes:
\[
\bar{\mathcal{T}}_{\pi_t}[\hat \rho]
= \hat \rho + \frac{1}{2}\,\big\langle [\hat X,[\hat X,\hat \rho]] \big\rangle_\phi + \mathcal{O}(|\alpha|^4),
\]
where the cubic term similarly vanishes. The quadratic term meanwhile is:
\begin{align*}
    \frac12\big\langle [\hat X,[\hat X,\hat \rho]]\big\rangle_\phi
    =|\alpha|^2\left(\mathcal D[\hat a]+\mathcal D[\hat a^\dagger]\right)\hat \rho .
\end{align*}
This means that:
\[
\bar{\mathcal{T}}_{\pi_t}[\hat \rho]
= \hat \rho + |\alpha|^2\left(\mathcal D[\hat a]+\mathcal D[\hat a^\dagger]\right)\hat \rho + \mathcal{O}(|\alpha|^4),
\]
which we recognize as the additive-noise channel $\bar{\mathcal{T}}_{\pi_t}\approx\mathcal{N}_\nu$ with $\nu=\abs{\alpha}^2$ up to error $\mathcal{O}[|\alpha|^4(\bar N+1)^2]$. This means that $\bar{\mathcal{E}}_{t_\text{int}}\approx\mathcal{C}_{\eta,\bar n_\text{th}}$ up to similar error, where then we identify:
\begin{align*}
    \bar n_\text{th} = \frac{\nu}{1-\eta} = \frac{|\alpha|^2}{1-\eta} = \frac{\frac{4\Omega_0^2}{\kappa^2}(1-\sqrt{\eta})^2}{1-\eta} = \frac{4\Omega_0^2}{\kappa^2}\tanh\left(\frac{\kappa t_\text{int}}{4}\right).
\end{align*}
For any $t_\text{int}$, i.e.\ even those comparable or larger than $1/\kappa$, we have that $\bar{\mathcal{E}}_{t_\text{int}}\approx\mathcal{C}_{\eta,\bar n_\text{th}}$ with $\bar n_\text{th}=\frac{4\Omega_0^2}{\kappa^2}\tanh\left(\frac{\kappa t_\text{int}}{4}\right)$ up to error $\mathcal{O}[|\alpha|^4(\bar N+1)^2]$. This means that we drop terms assuming that $\frac{4\Omega_0^2}{\kappa^2}(1-\sqrt{\eta})^2(\bar N+1)\ll1$. The resulting $\bar n_\text{th}$ for the thermal loss channel depends on time.

Let us now discuss subsequently taking the short interrogation time limit of $\kappa t_\text{int}\ll1$. In this limit, $\eta=e^{-\kappa t_\text{int}}\approx 1-\kappa t_\text{int}$ such that $\alpha=-i\frac{2\Omega_0}{\kappa}(1-\sqrt{\eta})e^{i\phi}\approx -i\Omega_0 t_\text{int}e^{i\phi}$ and $\abs{\alpha}\approx\Omega_0t_\text{int}$. The condition on dropping the error term is now that $\Omega_0^2t_\text{int}^2(\bar N+1)\ll1$. Similarly, the effective thermal occupation is then:
\begin{align*}
    \bar n_\text{th}
        =\frac{\Omega_0^2t_\text{int}}{\kappa}\left[1+\mathcal{O}(\kappa^2t_\text{int}^2)\right],
\end{align*}
which means that $\bar n_\text{th}\approx\frac{\Omega_0^2t_\text{int}}{\kappa}$, assuming that $\kappa t_\text{int}\ll1$. The resulting $\bar n_\text{th}$ still depends on time in this short time limit.

Hence, while the evolution for this first signal model corresponds to a thermal loss channel at a given time, this channel cannot be described as the solution of the master equation in Eq.~\ref{eq:me} for all times. The reason for this is that $\bar n_\text{th}$ for this channel depends on time $t_\text{int}$, while for the master equation $\bar n_\text{th}$ is constant.
The reason for this model not matching the master equation for fixed $\Omega_0$ is because the underlying signal is coherent such that the magnitude of the displacement (for the fixed-radius model) grows as $t_\text{int}$ rather than $\sqrt{t_\text{int}}$ as for a random walk. This latter case is what we examine in the second model below where the radius is no longer fixed.

If we defined $\gamma\approx\Omega_0^2t_\text{int}$ such that $\bar n_\text{th}=\frac{\gamma}{\kappa}$, then we have that $\gamma$ depends on time for a fixed $\Omega_0$. For $\gamma$ to be fixed for all time (which is what we assume in Eq.~\ref{eq:me}), we would instead have to require $\Omega_0$ to fall as $1/\sqrt{t_\text{int}}$ over time, which is a different model. This scaling effectively counters the coherent build-up of the displacement as the signal gets weaker over time.

\subsection{Stochastic signal as a complex-valued random process}
The second model of a stochastic signal in the main text is that $\Omega(t)$ is a Gaussian complex-valued random process. As discussed above, this implies that $\pi_t$ in Eq.~\ref{eq:pushforward measure} is a 2D Gaussian distribution for a given $t$, which ensures that the channel is Gaussian. We further assume that $\Omega(t)$ is an isotropic (rotationally invariant) random process. This implies that $\pi_{t}$ is then also isotropic and thus $\bar{\mathcal{T}}_{\pi_t}=\mathcal{N}_\nu$ for some $\nu$. We now calculate $\nu$ for the Lorentzian model.

From Eq.~\ref{eq:a solution to a}, we have that $\alpha_{t}=-i\int_{0}^{t}\text{d}s\;e^{-\kappa\left(t-s\right)/2}\Omega(s)$ such that $\langle \alpha_{t} \rangle=0$ given that $\langle \Omega(s)\rangle=0$ is a zero-mean process. We then have that:
\begin{align*}
    \langle |\alpha_{t}|^2 \rangle = e^{-\kappa t}\int_{0}^{t}\text{d}s\;\int_{0}^{t}\text{d}s'\;e^{\kappa(s+s')/2} \langle \Omega(s)\Omega^*(s') \rangle,\qquad
    \langle \alpha_{t}^2 \rangle = e^{-\kappa t}\int_{0}^{t}\text{d}s\;\int_{0}^{t}\text{d}s'\;e^{\kappa(s+s')/2} \langle \Omega(s)\Omega(s') \rangle.
\end{align*}
We assume that $\langle \Omega(s)\Omega(s') \rangle=0$ such that the complex-valued process is rotationally invariant, which implies that $\langle \alpha_{t}^2 \rangle =0$. We also assume that $\langle \Omega(s)\Omega^*(s') \rangle=\Omega_0^2e^{-\Gamma_\phi\abs{s-s'}}$ such that the process has correlation time $\tau_\phi=1/\Gamma_\phi$ and we have that:
\begin{align}\label{eq:nu for noise}
    \nu := \langle |\alpha_{t}|^2 \rangle = \frac{\Omega_{0}^{2}}{\kappa/2+\Gamma_{\phi}}\left[\frac{1-e^{-\kappa t}}{\kappa/2}-2e^{-\kappa t/2}\frac{e^{-\Gamma_{\phi}t}-e^{-\kappa t/2}}{\kappa/2-\Gamma_{\phi}}\right],
\end{align}
which defines the resulting isotropic noise $\nu$ as a function of time and shows that $\bar{\mathcal{T}}_{\pi_t}=\mathcal{N}_\nu$ for this value of $\nu$. Recall that $\nu=(1-\eta)\bar n_\text{th}$ such that this also defines $\bar n_\text{th}$ as a function of time.

We have shown that the solution for the random process for a given interrogation time $t_\text{int}$ is the thermal loss channel $\mathcal{C}_{\eta,\bar n_\text{th}}$, where $\eta=e^{-\kappa t}$ and $\bar n_\text{th}$ from Eq.~\ref{eq:nu for noise} generally depend on time. To show that it is the solution to the master equation in Eq.~\ref{eq:me} for some fixed $\gamma$ requires further assumptions however, namely that $t_\text{int}\gg\tau_\phi$ and $\kappa\tau_\phi\ll1$ (equivalently, $\Gamma_\phi t_\text{int}\gg1$ and $\kappa \ll \Gamma_{\phi}$). 
Starting from Eq.~\ref{eq:nu for noise} and using $t_\text{int}\gg\tau_\phi$ to drop the $e^{-\Gamma_{\phi}t}$ term, we have that:
\begin{align*}
    \nu \approx \frac{\Omega_{0}^{2}}{\kappa/2+\Gamma_{\phi}}\left[\frac{1-e^{-\kappa t}}{\kappa/2}+\frac{2e^{-\kappa t}}{\kappa/2-\Gamma_{\phi}}\right].
\end{align*}
We then use $\kappa \ll \Gamma_{\phi}$ to expand in series in the small parameter $\kappa\tau_\phi\ll1$: \begin{align*}
    \nu\approx (1-e^{-\kappa t})\frac{2\Omega_{0}^{2}}{\Gamma_{\phi}\kappa}-(1+e^{-\kappa t})\frac{\Omega_{0}^{2}}{\Gamma_{\phi}^2} + \mathcal{O}\left(\frac{\kappa\Omega_0^2}{\Gamma_{\phi}^3}\right).
\end{align*}
The first term is what we want. The ratio of the second, error term to the first term is $-\frac{\kappa }{2 \Gamma_ \phi }\coth(\frac{\kappa  t}{2})$. To drop the error term, we thus want $\frac{\kappa }{2 \Gamma_ \phi }\coth(\frac{\kappa  t_\text{int}}{2})\ll1$ to be small, which is true given $\kappa\tau_\phi\ll1$ if $\kappa t_\text{int}\sim1$ or $\kappa t_\text{int}\gg1$ and given $t\gg\tau_\phi$ if $\kappa t_\text{int}\ll1$. Therefore, in the limit of $t_\text{int}\gg\tau_\phi$ and $\kappa\tau_\phi\ll1$, the noise becomes:
\begin{align*}
    \nu\approx (1-e^{-\kappa t_\text{int}})\frac{2\Omega_{0}^{2}}{\Gamma_{\phi}\kappa}=(1-\eta)\bar n_\text{th},
\end{align*}
where $\eta=e^{-\kappa t_\text{int}}$ and $\bar n_\text{th}=\gamma/\kappa$ with $\gamma=2\Omega_{0}^{2}/\Gamma_{\phi}$ now time-independent. 
The added noise $\nu$ scales linearly with time for short times $\kappa t_\text{int}\ll1$ because of the diffusive assumption $t_\text{int}\gg\tau_\phi$, whereas for $t_\text{int}\ll\tau_\phi$ the scaling is instead quadratic. The white noise case is the short correlation time limit $\tau_\phi\to0$ for a fixed $\gamma$.

The same results for the second model above can alternatively be shown from the second-order cumulant expansion using the Born-Markov approximation as follows, similarly assuming that $\tau_\phi\ll\min(t_\text{int},1/\kappa)$. We assume that the Gaussian process is rotationally invariant ($\langle \Omega(t)\Omega(t-\tau)\rangle=0$) and stationary such that we define the time correlation function $C(\tau):=\langle{\Omega(t)\Omega^*(t-\tau)}\rangle$. We further assume that $C(\tau)$ is real-valued and an even function of $\tau$. We define the power spectral density as $ S_\Omega(\omega) := \int_{-\infty}^{\infty} \text{d}\tau\;e^{i\omega\tau} C(\tau)$ and will now prove that $\gamma:= S_\Omega(0)$ is the resulting diffusion rate. For the Lorentzian signal process $C(\tau)=\Omega_0^2e^{-\Gamma_\phi\abs{\tau}}$ we have $S_\Omega(\omega)=\frac{\gamma}{1+(\omega/\Gamma_\phi)^2}$ and $\gamma=\frac{2\Omega_0^2}{\Gamma_\phi}$. This recovers the white noise limit of $C(\tau)=\gamma\delta(\tau)$ and $S_\Omega(\omega)=\gamma$ if $\Gamma_\phi\to\infty$ is taken while $\gamma$ is fixed. 

The noise-averaged density matrix $\bar\rho(t) = \langle \rho(t)\rangle$ obeys the second-order cumulant master equation~\cite{rivas2012open,lidar2019lecture}:
\begin{equation}
  \deriv{\bar\rho}{t}
  = -\frac{1}{\hbar^2}\int_0^\infty \text{d}\tau\;
      \bigl\langle [H(t),[H(t-\tau),\bar\rho(t)]]\bigr\rangle,
  \label{eq:born_markov_classical}
\end{equation}
where we have used the Born--Markov approximation and stationarity of the noise to extend the upper integration limit to infinity. 
We also focus on the $\kappa=0$ case here.
Substituting the Hamiltonian from Eq.~\ref{eq:Hamiltonian}, we find that the integrand in Eq.~\ref{eq:born_markov_classical} becomes:
\begin{align*}
    \frac{1}{\hbar^2}\bigl\langle [H(t),[H(t-\tau),\bar\rho(t)]]\bigr\rangle
  &= \langle\Omega^*(t)\Omega^*(t-\tau)\rangle[a,[a,\bar\rho]]
   + \langle\Omega^*(t)\Omega(t-\tau)\rangle[a,[a^\dagger,\bar\rho]]
   \nonumber \\
  &\quad
   + \langle\Omega(t)\Omega^*(t-\tau)\rangle[a^\dagger,[a,\bar\rho]]
   + \langle\Omega(t)\Omega(t-\tau)\rangle[a^\dagger,[a^\dagger,\bar\rho]],
\end{align*}
where then using rotational invariance and the assumption that $C(\tau)$ is real and even, we get that:
\begin{align*}
    \frac{1}{\hbar^2}\bigl\langle [H(t),[H(t-\tau),\bar\rho(t)]]\bigr\rangle
    &= C(\tau)\left([a,[a^\dagger,\bar\rho]]
   +[a^\dagger,[a,\bar\rho]]\right).
\end{align*}
The remaining nested commutators can be expressed using Lindblad dissipators via a straightforward calculation which gives:
\begin{align*}
    \frac{1}{\hbar^2}\bigl\langle [H(t),[H(t-\tau),\bar\rho(t)]]\bigr\rangle
  = -2 C(\tau)
     \bigl(\mathcal D[a]\bar\rho + \mathcal D[a^\dagger]\bar\rho\bigr).
\end{align*}
Substituting into Eq.~\ref{eq:born_markov_classical}, we obtain the diffusion master equation for a $\gamma$ rate which we define as follows:
\begin{equation}  \label{eq:kappa_general}
  \deriv{\bar\rho}{t} = \gamma \left(\mathcal D[a]\bar\rho + \mathcal D[a^\dagger]\bar\rho\right), \qquad \gamma := 2\int_0^\infty \text{d}\tau\, C(\tau).
\end{equation}
Informally, if the Born--Markov approximation holds, adding back the loss $\kappa$ from Eq.~\ref{eq:Hamiltonian} to this diffusion equation yields the master equation in Eq.~\ref{eq:me}. We have not proved this formally here in this second approach. However, we have shown this in the first derivation in Eq.~\ref{eq:nu for noise}.

\newpage
\section{Optimality of TMSV for a given energy: Proof of Claim~\ref{claim:ECQFI and TMSV}}
\label{supp_sec:tmsv_qfi}
We now prove that TMSV is generically optimal for estimating a single parameter of the thermal loss channel with the other degree-of-freedom fixed. We first calculate the generic TMSV QFI and then show that it attains an upper bound on the ECQFI.

\subsection{TMSV QFI}
In general, the TMSV QFI can be calculated using the following multi-mode Gaussian QFI formula from Ref.~\cite{monras2013phase}:
\begin{align}
\IQ=2\langle\partial_{\theta}\Sigma|\left(4\Sigma\otimes\Sigma-J\otimes J\right)^{-1}|\partial_{\theta}\Sigma\rangle,
\label{supp_eq:gaussian_qfi_general}
\end{align}
where $\Sigma$ is the covariance matrix of the TMSV after the thermal channel evolution and $J$ is the symplectic matrix.
Given a $\left(Q_{1},Q_{2},P_{1},P_{2}\right)$
ordering of the quadratures where  mode 1 is the ancillary mode, then $\Sigma,$ $J$ are given by: \begin{align}
\Sigma=\left(\begin{array}{cccc}
\bar{N}+\frac{1}{2} & \sqrt{\eta\bar{N}(\bar{N}+1)} & 0 & 0\\
\sqrt{\eta\bar{N}(\bar{N}+1)} & \eta\bar{N}+\left(1-\eta\right)\bar{n}_{\text{th}}+\frac{1}{2} & 0 & 0\\
0 & 0 & \bar{N}+\frac{1}{2} & -\sqrt{\eta\bar{N}(\bar{N}+1)}\\
0 & 0 & -\sqrt{\eta\bar{N}(\bar{N}+1)} & \eta\bar{N}+\left(1-\eta\right)\bar{n}_{\text{th}}+\frac{1}{2}
\end{array}\right),\quad
    J=\left(\begin{array}{cccc}
0 & 0 & 1 & 0\\
0 & 0 & 0 & 1\\
-1 & 0 & 0 & 0\\
0 & -1 & 0 & 0
\end{array}\right).
\end{align}
Here $\eta=e^{-\left( \Gamma-\gamma \right)t}$ is the transmissivity of the thermal channel, and $\bar{n}_{\text{th}}=\frac{\gamma}{\Gamma-\gamma}$ is the thermal occupation of this thermal channel.

For the thermal loss channel $C_{\eta,\bar n_\text{th}}$ where $\eta$ and $\bar n_\text{th}$ are implicitly functions of some generic real parameter $\theta$, the TMSV QFI with respect to $\theta$ in Eq.~\ref{supp_eq:gaussian_qfi_general} is, after some simplification in \texttt{Mathematica}:
\begin{align}\label{eq:TMSV QFI}
    \IQ = \frac{\bar n_{\text{th}} \left(\bar n_{\text{th}}+1\right) \dot \eta ^2 \left(\eta  \left((2 \bar N+1) \bar n_{\text{th}}-\bar N^2\right)+\bar N (\bar N+1)\right)+2 (2 \bar N+1) (\eta -1) \eta  \bar n_{\text{th}} \left(\bar n_{\text{th}}+1\right) \dot \eta  \dot{\bar n}_{\text{th}}+(\eta -1)^2 \eta  \left((2 \bar N+1) \bar n_{\text{th}}+\bar N+1\right) \dot{\bar n}_{\text{th}}{}^2}{(\eta -1) \eta  \bar n_{\text{th}} \left(\bar n_{\text{th}}+1\right) \left((2 \bar N+1) (\eta -1) \bar n_{\text{th}}+\bar N \eta -\bar N-1\right)}.
\end{align}
Let us define the coefficients $c_j$ such that the QFI above is:
\begin{align*}
    \IQ = \frac{c_1 \bar N^2 + c_2 \bar N + c_3}{c_4 \bar N + c_5},
\end{align*}
where the coefficients are given as follows:
\begin{align}
    c_1 &= (1-\eta ) \bar n_{\text{th}} (\bar n_{\text{th}}+1)\dot{\eta }^2,  \\
    c_2 &= \bar n_{\text{th}} (\bar n_{\text{th}}+1) (2 \eta  \bar n_{\text{th}}+1)\dot{\eta }^2 
        \nonumber\\&\quad- 4 (1-\eta) \eta  \bar n_{\text{th}} (\bar n_{\text{th}}+1) \dot{\eta } \dot{\bar n}_{\text{th}}
        \nonumber\\&\quad+(1-\eta)^2 \eta  (2 \bar n_{\text{th}}+1) \dot{\bar n}_{\text{th}}^2, \nonumber\\
    c_3 &= \eta  (\bar n_{\text{th}}+1) [\bar n_{\text{th}}\dot{\eta } -(1-\eta) \dot{\bar n}_{\text{th}}]^2, \nonumber\\
    c_4 &= (1-\eta)^2 \eta \bar n_{\text{th}} (\bar n_{\text{th}}+1) (2 \bar n_{\text{th}}+1), \nonumber\\
    c_5 &= (1-\eta) \eta \bar n_{\text{th}} (\bar n_{\text{th}}+1) [(1-\eta) \bar n_{\text{th}}+1].\nonumber
\end{align}
This is the result that we give in the End Matter. Direct computation and simplification of these coefficients leads to Eqs.~\ref{eq:noise sensing},~\ref{eq:loss estimation},~\ref{eq:Scenario 1 coefficients} in the manuscript.
We now need to show that this QFI is optimal for a given $\bar N$.

\newpage
\subsection{Upper bound on the ECQFI}
\label{supp_sec:hiding_unitary_ECQFI}
To find an upper bound on the ECQFI, we first decompose the thermal loss channel $\mathcal{C}_{\eta, \bar n_\text{th}}$ into a loss channel into vacuum with transmissivity $\eta'=\eta/G$ followed by a quantum-limited amplifier with gain $G=1+(1-\eta)\bar n_\text{th}$. One possible representation of this channel $\mathcal{C}_{\eta, \bar n_\text{th}}(\hat\rho)=\text{Tr}_{\hat b\hat c}[\hat U_{\text{amp}}\hat U_\text{bs}\hat\rho\hat U_\text{bs}^\dagger\hat U_{\text{amp}}^\dagger]$ is as a beamsplitter $\hat U_\text{bs}=\exp[\xi_{1}(\hat a^\dagger \hat b-\hat a \hat b^{\dagger})]$ with $\cos^2(\xi_1)=\eta'$, followed by parametric driving $\hat U_{\text{amp}}=\exp(\xi_{2}[\hat a^\dagger \hat c^\dagger-\hat a\hat  c)]$ with $\cosh^2(\xi_2)=G$, and then tracing out of the two ancilla modes $\hat b$ and $\hat c$. Here, $\xi_1$ and $\xi_2$ both implicitly depend on the parameter $\theta$ through $\eta$ and $\bar n_\text{th}$ as:
\begin{align}\label{eq:xi12}
    \xi_{1}:=\arccos\left(\sqrt{\frac{\eta}{1+(1-\eta)\bar n_\text{th}}}\right), \quad \xi_{2}:=\text{arccosh}\left(\sqrt{1+(1-\eta)\bar n_\text{th}}\right).
\end{align}
The Hamiltonian of $\hat U_\text{bs}=\exp(-i\xi_1 \hat H_\text{bs})$ is $\hat H_\text{bs}=i(\hat a^\dagger \hat b - \hat a\hat b^\dagger)$ and of $\hat U_\text{amp}=\exp(-i\xi_2 \hat H_\text{amp})$ is $\hat H_\text{amp}=i(\hat a^\dagger \hat c^\dagger - \hat a\hat c)$.

The QFI of the pure state $\hat U_{\text{amp}}\hat U_\text{bs}\ket{\psi}$ with $\hat\rho=\ket{\psi}\bra{\psi}$ before tracing out the ancillae is an upper bound on the achievable QFI of $\mathcal{C}_{\eta, \bar n_\text{th}}(\hat\rho)$ thereafter, due to the QFI not increasing under parameter-invariant channels. Furthermore, we can introduce a hiding unitary $\hat U_\text{hide}$ which depends on the parameter $\theta$ but acts only on the ancillae such that it does not affect the state after tracing out the ancilla. The QFI of $\hat U_\text{hide}\hat U_{\text{amp}}\hat U_\text{bs}\ket{\psi}$ is therefore still an upper bound on the QFI $\IQ$ of $\mathcal{C}_{\eta, \bar n_\text{th}}(\hat\rho)$, and a good choice of hiding unitary can make this upper bound tight~\cite{latune2013quantum,gardner2025stochastic}. (Naturally, a poor choice can instead loosen the upper bound, so we need to be careful in our selection.) We choose $\hat U_\text{hide}=\exp[g\theta(\hat b^\dagger \hat c^\dagger- \hat b\hat c)]$ where $g\in\mathbb{R}$ is a control parameter to be minimized over. The Hamiltonian of $\hat U_\text{hide}=\exp(-i\theta \hat H_\text{hide})$ is thus $\hat H_\text{hide}=ig(\hat b^\dagger \hat c^\dagger- \hat b\hat c)$.
This choice was inspired from terms like $\hat b\hat c$ appearing in the Hamiltonian $-i\hat U_\text{bs}^\dagger\hat U_{\text{amp}}^\dagger\partial_{\theta}(\hat U_{\text{amp}}\hat U_\text{bs})$ that we wanted to cancel.
With the hiding unitary included, the overall Hamiltonian of $\hat U:=\hat U_\text{hide}\hat U_{\text{amp}}\hat U_\text{bs}$ is thus $\hat H:=-i\hat U^\dagger\partial_{\theta}\hat U$ and the QFI upper bound is $4\,\text{Var}[\hat H]$. This Hamiltonian simplifies to:
\begin{align*}
    \hat H
    &=- \dot \xi_1 \hat H_\text{bs}-\dot \xi_2\hat U_{\text{bs}}^\dagger \hat H_\text{amp} \hat U_{\text{bs}} -\hat U_{\text{bs}}^\dagger \hat U_\text{amp}^\dagger \hat  H_\text{hide} \hat U_\text{amp} \hat U_{\text{bs}}
    \\&= i g \left(\cosh (\xi_2) \left(\sin (\xi_1) \left(\hat{c}^{\dagger } \hat{a}^{\dagger }-\hat{a} \hat{c}\right)+\cos (\xi_1) \left(\hat{b} \hat{c}-\hat{c}^{\dagger } \hat{b}^{\dagger }\right)\right)+\sinh (\xi_2) \left(\hat{a}^{\dagger } \hat{b}-\hat{b}^{\dagger } \hat{a}\right)\right)
    \\&+i \left(\dot{\xi}_2 \left(\cos (\xi_1) \left(\hat{a} \hat{c}-\hat{c}^{\dagger } \hat{a}^{\dagger }\right)+\sin (\xi_1) \left(\hat{b} \hat{c}-\hat{c}^{\dagger } \hat{b}^{\dagger }\right)\right)+\dot{\xi}_1 \left(\hat{b}^{\dagger } \hat{a}-\hat{a}^{\dagger } \hat{b}\right)\right)
\end{align*}
such that the upper bound as a function of $g$ is:
\begin{align*}
    4\,\text{Var}[\hat H] &= g^2 \left(4 \cosh ^2(\xi_2)+4 \bar N  \left(\sin ^2(\xi_1) \cosh ^2(\xi_2)+\sinh ^2(\xi_2)\right)\right)
    \\&-8 g \bar N  \left(\dot{\xi}_1 \sinh (\xi_2)+\dot{\xi}_2 \sin (\xi_1) \cos (\xi_1) \cosh (\xi_2)\right)
    \\&+4 \dot{\xi}_2^2+4 \bar N  \left(\dot{\xi}_2^2 \cos ^2(\xi_1)+\dot{\xi}_1^2\right)
\end{align*}
which minimized over $g$ yields:
\begin{align}
   \min_g 4\,\text{Var}[\hat H] &= \frac{4 \dot{\xi}_2^2 \cosh ^2(\xi_2)+4 \bar N^2 \left(\dot{\xi}_1 \sin (\xi_1) \cosh (\xi_2)-\dot{\xi}_2 \cos (\xi_1) \sinh (\xi_2)\right)^2+4 \bar N \left(\dot{\xi}_1^2 \cosh ^2(\xi_2)+\dot{\xi}_2^2 \cosh (2 \xi_2)\right)}
   {\cosh ^2(\xi_2)+\bar N \left(\sin ^2(\xi_1) \cosh ^2(\xi_2)+\sinh ^2(\xi_2)\right)}
\label{supp_eq:ECQFI_general_expression}   
\end{align}
We now want to express this in terms of parameter derivative of $\eta$ and $\bar n_\text{th}$ using Eq.~\ref{eq:xi12}:
\begin{align*}
                \dot{\xi}_1 &= \frac{-\left(\left(\bar n_{\text{th}}+1\right) \dot\eta \right)-(\eta -1) \eta  \dot{\bar n}_{\text{th}}}{2 \sqrt{-\frac{(\eta -1) \eta  \left(\bar n_{\text{th}}+1\right)}{\left((\eta -1) \bar n_{\text{th}}-1\right){}^2}} \left((\eta -1) \bar n_{\text{th}}-1\right){}^2}
    \\\dot{\xi}_2 &= \frac{-\bar n_{\text{th}} \dot\eta-(\eta -1) \dot{\bar n}_{\text{th}}}{2 \sqrt{1-(\eta -1) \bar n_{\text{th}}} \sqrt{\sqrt{1-(\eta -1) \bar n_{\text{th}}}-1} \sqrt{\sqrt{1-(\eta -1) \bar n_{\text{th}}}+1}}
        \end{align*}
such that Eq.~\ref{supp_eq:ECQFI_general_expression} becomes, after some simplification in \texttt{Mathematica}:
\[
\resizebox{\textwidth}{!}{$
\begin{aligned}
    \min_g 4\text{Var}[\hat H] &= \frac{\bar n_{\text{th}} \left(\bar n_{\text{th}}+1\right) \dot\eta ^2 \left((2 \bar N+1) \eta  \bar n_{\text{th}}+\bar N (\bar N (-\eta )+\bar N+1)\right)+2 (2 \bar N+1) (\eta -1) \eta  \bar n_{\text{th}} \left(\bar n_{\text{th}}+1\right) \dot\eta \dot{\bar n}_{\text{th}}+(\eta -1)^2 \eta  \left((2 \bar N+1) \bar n_{\text{th}}+\bar N+1\right) \dot{\bar n}_{\text{th}}{}^2}{(\eta -1) \eta  \bar n_{\text{th}} \left(\bar n_{\text{th}}+1\right) \left((2 \bar N+1) (\eta -1) \bar n_{\text{th}}+\bar N \eta -\bar N-1\right)}
\end{aligned}
$}
\]
which matches the TMSV QFI in Eq.~\ref{eq:TMSV QFI} for any parameter $\theta$ and any $\bar N$, completing the proof.

While TMSV is optimal for all cases, whether an entanglement gap exists changes dramatically between cases. A generic understanding of the entanglement gap and also simultaneous estimation of $\eta$ and $\bar n_\text{th}$ would be interesting for future work.

\newpage
\section{Noise sensing SNR}
\label{sec:noise_sensing_snr}
We now derive the TOP SNR for noise sensing (estimating $\sqrt{\gamma}$ with $\kappa$ fixed) stated in the main text.
The ECQFI in Eq.~\ref{eq:noise sensing} is:
\begin{align*}
    \IQ^\text{E}
    =
    \frac{
    4(1-\eta)[(2\gamma+\kappa)\bar N+\gamma+\kappa]
    }{
    (\gamma+\kappa)[(1-\eta)\{(2\gamma+\kappa)\bar N+\gamma\}+\kappa]
    } .
    \end{align*}
In the weak-signal limit $\gamma\ll\kappa$, this
reduces to:
\begin{align*}
    \IQ^\text{E}
    =
    \frac{4(1-e^{-\kappa t_\text{int}})(\bar N+1)}
    {\kappa[(1-e^{-\kappa t_\text{int}})\bar N+1]}
    +\mathcal{O}(\gamma/\kappa).
    \end{align*}
We take the short-interrogation-time limit
$\kappa t_\text{int}\ll1$. We will show below that this assumption is consistent in the high energy limit as we have that $\kappa t_\text{int}^{*}\ll1$ for a given dead time $\tau_\text{dead}$ and loss $\kappa$. Then, the ECQFI becomes:
\begin{align*}
    \IQ^\text{E}
    \approx
    \frac{4t_\text{int}(\bar N+1)}{1+\kappa\bar N t_\text{int}}.
    \end{align*}
The SNR squared is therefore:
\begin{align*}
    \left(\text{SNR}_Q^\text{E}\right)^2
    = \frac{\gamma T \IQ}{t_\text{int}+\tau_\text{dead}}
    \approx
    4\gamma T(\bar N+1)
    \frac{t_\text{int}}{(t_\text{int}+\tau_\text{dead})(1+\kappa\bar N t_\text{int})}.
    \end{align*}
Maximizing over $t_\text{int}$ gives:
\begin{align*}
    t_\text{int}^*
    =
    \sqrt{\frac{\tau_\text{dead}}{\kappa\bar N}},
    \end{align*}
and hence:
\begin{align*}
    \left(\text{SNR}_Q^{E,\text{TOP}}\right)^2
    \approx
    \frac{
    4\gamma T(\bar N+1)
    }{
    \left(1+\sqrt{\kappa\bar N\tau_\text{dead}}\right)^2
    } .
    \end{align*}
This is Eq.~\ref{eq:TOP noise sensing} of the main text. This implies that the interrogation time is short $\kappa t_\text{int}^*\ll1$ if $\kappa \tau_\text{dead}\ll\bar N$ which is always possible in the high energy limit for a given $\kappa \tau_\text{dead}$. This means that our assumption that $\kappa t_\text{int}\ll1$ is consistent.

For short dead times
$\kappa\tau_\text{dead}\ll\min(1/\bar N, \bar N)$, the TOP SNR squared is:
\begin{align*}
    \left(\text{SNR}_Q^{E,\text{TOP}}\right)^2
    \approx
    \left(1-2\sqrt{\kappa\bar N\tau_\text{dead}}\right)
    4\gamma T(\bar N+1).
    \end{align*}
Whereas, for long dead times $1/\bar N\ll\kappa\tau_\text{dead}\ll \bar N$ and high energies $\bar N\gg1$, the TOP SNR squared is:
\begin{align}
    \left(\text{SNR}_Q^{E,\text{TOP}}\right)^2
    \approx
    \frac{4\gamma T}{\kappa\tau_\text{dead}}.
    \label{eq:noise_long_dead_time}
\end{align}
This proves the results in the main text.

\newpage
\section{Asymmetric noise sensing}
Let us briefly discuss the asymmetric case of noise sensing as an example for when TMSV is not optimal at finite energy. Recall that for the symmetric (isotropic) case, the thermal loss channel acts on a single-mode Gaussian state as $\vec\mu\to\sqrt{\eta}\vec\mu$ and $\Sigma\to\eta\Sigma+(1-\eta)\Sigma_\text{th}$, where $\Sigma_\text{th}=(\frac{1}{2}+\bar n_\text{th})\,\text{diag}(1,1)$. In the fully asymmetric case, the channel instead acts as $\vec\mu\to\sqrt{\eta}\vec\mu$ and $\Sigma\to\eta\Sigma+(1-\eta)\,\text{diag}(\frac{1}{2},\frac{1}{2}+\varsigma^2)$, which is the same as $\Sigma\to\eta\Sigma+(1-\eta)\Sigma_\text{vac}+\,\text{diag}(0,\sigma^2)$ where $\Sigma_\text{vac}=\Sigma_\text{th}|_{\bar n_\text{th}=0}$ and $\sigma^2=(1-\eta)\varsigma^2$. This is the channel $\bar{\mathcal{T}}_{p}\circ\mathcal{L}_{\eta}$, loss followed by asymmetric additive noise, where $p$ is the zero-mean Gaussian with covariance matrix $\text{diag}(0,\sigma^2)$. We consider here the case without classical noise; the case of classical noise with covariance matrix $\text{diag}(\sigma_x^2,\sigma_p^2+\sigma^2)$ is discussed in Ref.~\cite{gardner2025stochastic}.

\begin{figure}
    \centering
    \includegraphics[width=0.7\linewidth]{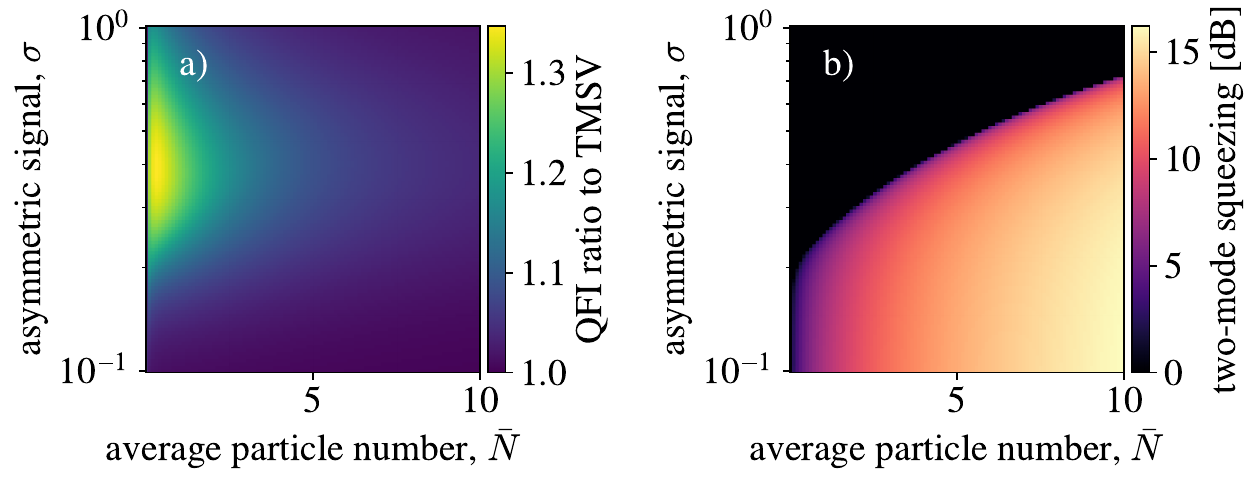}
    \caption{\textbf{a)} For the asymmetric thermal loss channel, gain in QFI compared to the TMSV QFI from applying some amount of single-mode squeezing with parameter $s$ to the TMSV state. We adjust the two-mode squeezing parameter $r$ to maintain the average particle number $\bar N$ in the signal mode. The gain is higher for $\eta$ closer to unity; this plot is for $\eta=0.9$. \textbf{b)} Optimal two-mode squeezing parameter $r$ showing the transition from TMSV (large $r$) to SMSV ($r=0$) as $\sigma$ increases for a fixed $\bar N$. Here, we quote the amount of two-mode squeezing in decibels as $r_\text{dB}=10\log_\text{10}(e^{2r})=20\,r\log_\text{10}(e)$ where $r$ is the two-mode squeezing parameter.}
    \label{fig:asymmetric TMSV}
\end{figure}

Let us consider applying single-mode squeezing with parameter $s$ to one half of a TMSV state with two-mode squeezing parameter $r$ such that the covariance matrix in the $(\hat x_1,\hat p_1, \hat x_2,\hat p_2)^\mathsf{T}$ basis becomes:
\begin{align*}
    \Sigma = \left(
\begin{array}{cccc}
 \frac{1}{2} e^{2 s} \cosh (2 r) & 0 & \frac{1}{2} e^s \sinh (2 r) & 0 \\
 0 & \frac{1}{2} e^{-2 s} \cosh (2 r) & 0 & -\frac{1}{2} e^{-s} \sinh (2 r) \\
 \frac{1}{2} e^s \sinh (2 r) & 0 & \frac{1}{2} \cosh (2 r) & 0 \\
 0 & -\frac{1}{2} e^{-s} \sinh (2 r) & 0 & \frac{1}{2} \cosh (2 r) \\
\end{array}
\right).
\end{align*}
The average number of particles in the first mode is now $\bar N = \frac{1}{2} (\cosh (2 r) \cosh (2 s)-1)$ whereas the second mode still has $\sinh^2(r)$ particles. Supposing that the first mode then passes through the asymmetric thermal loss channel above and the second mode experiences nothing, the covariance matrix then becomes:
\begin{align*}
    \Sigma' = \left(
\begin{array}{cccc}
 \frac{1-\eta }{2}+\frac{1}{2} \eta  e^{2 s} \cosh (2 r) & 0 & \frac{1}{2} \sqrt{\eta } e^s \sinh (2 r) & 0 \\
 0 & \frac{1-\eta }{2}+\frac{1}{2} \eta  e^{-2 s} \cosh (2 r)+\sigma ^2 & 0 & -\frac{1}{2} \sqrt{\eta } e^{-s} \sinh (2 r) \\
 \frac{1}{2} \sqrt{\eta } e^s \sinh (2 r) & 0 & \frac{1}{2} \cosh (2 r) & 0 \\
 0 & -\frac{1}{2} \sqrt{\eta } e^{-s} \sinh (2 r) & 0 & \frac{1}{2} \cosh (2 r) \\
\end{array}
\right).
\end{align*}
The QFI for this zero-mean state with respect to $\sigma$ is verbose and calculated from the standard Gaussian QFI formula as~\cite{monras2013phase}:
\[
\IQ(\sigma)=
\frac{
\begin{aligned}
&e^{-4s}\Big[
32\eta^2 e^{8s}\sigma^2\cosh^2(2r)
+(\eta-1)e^{4s}
\Big(
5\eta^2+14(\eta-1)\sigma^2-2\eta \\
&\quad
-4\big(\eta^2-4(\eta-1)\sigma^2+1\big)\cosh(4r)
-(\eta-1)(\eta-2\sigma^2-1)\cosh(8r)
+5
\Big) \\
&\quad
+4\eta(\eta-1)e^{6s}\cosh(2r)
\Big(
-\eta+(\eta-6\sigma^2-1)\cosh(4r)-10\sigma^2+1
\Big) \\
&\quad
+4\eta(\eta-1)^2 e^{2s}\sinh(2r)\sinh(4r)
\Big]
\end{aligned}
}{
\begin{aligned}
&\Big[
3\eta^2-3(\eta-1)\sigma^2-2\eta
+(\eta-\sigma^2-1)
\big((\eta-1)\cosh(4r)-4\eta\cosh(2r)\cosh(2s)\big) \\
&\quad
+4\eta\sigma^2\cosh(2r)\sinh(2s)+3
\Big] \\
&\quad\times
\Big[
(\eta-1)(3\eta-2\sigma^2+1)
+(\eta-1)(\eta-2\sigma^2-1)\cosh(4r) \\
&\quad
+4\eta\cosh(2r)
\big((-\eta+\sigma^2+1)\cosh(2s)+\sigma^2\sinh(2s)\big)
\Big]
\end{aligned}
}
\]
In the high energy limit $r\to\infty$, for any fixed $s$ (including $s=0$ for TMSV), the result is $\IQ(\sigma)=2/(1-\eta+\sigma^2)$ which achieves the high energy ECQFI~\cite{gardner2025stochastic}. For a fixed and finite energy $\bar N$, however, a nonzero $s$ value may offer an advantage over the $s=0$ standard TMSV with the same energy. We emphasize that we compare the QFI for different $s$ values with energy $\bar N = \frac{1}{2} (\cosh (2 r) \cosh (2 s)-1)$ fixed such that the $r$ value changes accordingly. We show in Fig.~\ref{fig:asymmetric TMSV} numerically that an advantage of roughly 30\% in the QFI (in power units) can be achieved for certain values of $\eta$, $\bar{N}$, and $\sigma$ from using some amount of additional single-mode squeezing or even by fully switching over from TMSV to single-mode squeezed vacuum (SMSV).
For example, for $\eta=0.9$ and $\sigma=0.3$ and $\bar N=0.5$ a gain of $31\%$ is achieved for $s\approx0.66$ which is the maximum value of $s$, indicating complete conversion to SMSV. For a fixed $\bar N$, we see a fairly sharp transition in $r$ from positive to zero as $\sigma$ increases and the optimal state within this manifold changes from standard TMSV, to some intermediate state, and finally to SMSV. 
The main gain is thus achieved for relatively small $\bar N$ (since for large $\bar N$ the TMSV is already optimal), and intermediate values of $\sigma$.
The intuition is that as the signal grows comparable to the added vacuum noise from the loss, then the signal asymmetry matters and for finite energy we want to devote resources to only the quadrature which contains the signal.
The signal strength, however, cannot be too large. For sufficiently large $\sigma$, the QFI is limited by the classical $2/\sigma^2$ scaling, and any gain from input squeezing becomes marginal.
Note that in the lossless case, SMSV beats TMSV by a factor of two at the same energy~\cite{gardner2025stochastic}. This factor of two improvement only appears, however, in the lossless case which is not of experimental interest.

The key point here is that although TMSV eventually achieves the high energy limit asymptotically, there is an advantage at finite energy for using different states. Experimentally, however, we are most interested in the vanishingly small $\sigma$ regime, where TMSV is close to optimal at finite energy.
We leave to future work determining the optimal Gaussian state, optimal state overall, and the ECQFI for the asymmetric noise channel at finite energy.

\newpage
\section{Loss estimation SNR}
We now provide some additional details about the short and long dead time limits for loss estimation of $\sqrt\kappa$ for fixed $\bar n_\text{th}$.
The high energy limit of Eq.~\ref{eq:loss estimation} is:
\begin{align*}
    \IQ\approx\frac{4\kappa\eta t_\text{int}^2\bar N}{(1-\eta)\left(2 \bar n_{\text{th}}+1\right)},
\end{align*}
which holds for $(1-\eta)\bar N\gg2 \bar n_{\text{th}}+1$ because that is when we have $c_1 \bar N\gg c_2$ and $c_4\bar N\gg c_5$. 
Using the high energy QFI, the stationary points of $\frac{\IQ}{t_\text{int}+\tau_\text{dead}}$ occur at:
\begin{align}\label{eq:dead time illumination}
    \tau_\text{dead} =-\frac{t_\text{int}^* \left(e^{\kappa  t_\text{int}^*} (\kappa  t_\text{int}^*-1)+1\right)}{e^{\kappa  t_\text{int}^*} (\kappa  t_\text{int}^*-2)+2}.
\end{align}
In the short dead time limit of $\kappa \tau_\text{dead}\ll1$, this becomes simply $\tau _\text{dead}\approx\frac{\kappa  (t_\text{int}^*)^2}{2}$ such that the optimal interrogation time is $t^*_\text{int}\approx\sqrt{2\tau_\text{dead}/\kappa}$. The high energy condition of $(1-\eta)\bar N\gg2 \bar n_{\text{th}}+1$ then becomes $\kappa t \bar N\gg2 \bar n_{\text{th}}+1$. This implies that we need $2\kappa \tau _\text{dead}\gg[(2 \bar n_{\text{th}}+1)/\bar N]^2$, so there is some competition between the short dead time and high energy limits that we need to be careful of.
The SNR squared is:
\begin{align}
    (\text{SNR}_Q^{\text{E},\text{TOP}})^2
    = \frac{\kappa\IQ T}{t_\text{int}^*+\tau_\text{dead}}
        \approx \frac{\eta \kappa^2(t_\text{int}^*)^2}{(\kappa t_\text{int}^*+\kappa\tau_\text{dead})(1-\eta)}\frac{4\kappa T\bar N}{\left(2 \bar n_{\text{th}}+1\right)},
\end{align}
where the time-dependent factor is $\frac{\eta \kappa^2(t_\text{int}^*)^2}{(\kappa t_\text{int}^*+\kappa\tau_\text{dead})(1-\eta)}$. The expansion of $\frac{\eta \kappa^2(t_\text{int}^*)^2}{(1-\eta)}$ in $\kappa t_\text{int}^*$ is:
\begin{align*}
    \frac{\eta \kappa^2(t_\text{int}^*)^2}{(1-\eta)}=\kappa t_\text{int}^*-\frac{(\kappa t_\text{int}^*)^2}{2}+\mathcal{O}[(\kappa t_\text{int}^*)^3].    
\end{align*}
This means that at $\kappa t^*_\text{int}\approx\sqrt{2\kappa \tau_\text{dead}}$, then:
\begin{align*}
    \frac{\eta \kappa^2(t_\text{int}^*)^2}{(\kappa t_\text{int}^*+\kappa\tau_\text{dead})(1-\eta)}
    \approx\frac{\kappa t_\text{int}^*-\frac{(\kappa t_\text{int}^*)^2}{2}}{(\kappa t_\text{int}^*+\kappa\tau_\text{dead})}
    \approx \frac{\sqrt{2\kappa \tau_\text{dead}}-\kappa \tau_\text{dead}}{\sqrt{2\kappa \tau_\text{dead}}+\kappa \tau_\text{dead}}
    \approx1-\sqrt{2\kappa \tau_\text{dead}}.
\end{align*}
Adding back the time-independent factor of $\frac{4\kappa T\bar N}{\left(2 \bar n_{\text{th}}+1\right)}$ from above, this proves Eq.~\ref{eq:illumination, harsh approx} for the short dead time limit.

In the long dead time limit $\kappa \tau_\text{dead}\gg1$, then we want $\tau_\text{dead}$ in Eq.~\ref{eq:dead time illumination} to diverge. To make the denominator vanish, then we need to solve $e^{\kappa  t_\text{int}^*} (\kappa  t_\text{int}^*-2)+2=0$. This is solved by $t_{\text{int}}^*=\left[2+W(-2/e^{2})\right]/\kappa\approx1.59/\kappa$. The TOP SNR squared is then:
\begin{align}
    (\text{SNR}_Q^{\text{E},\text{TOP}})^2
    \approx \frac{\eta \kappa^2(t_\text{int}^*)^2}{(\kappa t_\text{int}^*+\kappa\tau_\text{dead})(1-\eta)}\frac{4\kappa T\bar N}{\left(2 \bar n_{\text{th}}+1\right)} 
    \approx \frac{\eta \kappa^2(t_\text{int}^*)^2}{\kappa\tau_\text{dead}(1-\eta)}\frac{4\kappa T\bar N}{\left(2 \bar n_{\text{th}}+1\right)} 
    \approx \frac{2.6\,T\bar N}{\tau_\text{dead}\left(2 \bar n_{\text{th}}+1\right)}
    ,
\end{align}
as we state in the main text.

\newpage
\section{Coherent state QFI: Independent gain estimation}
In the main text, we say that, for long dead times, the coherent state TOP SNR is $\sqrt{2.17\bar{N}\gamma^2 T/(\Gamma^{2}\tau_{\text{dead}})}$ for $t_{\text{int}}^*=\left[1+\sqrt{1+\Gamma/(\gamma\bar{N})}\right]/\Gamma$, given $\gamma \ll \Gamma$ and $\bar N\gg1$. We now prove this result.

The coherent state QFI with respect to $\sqrt \gamma$ for fixed $\Gamma$ is (using the Gaussian QFI formula from Ref.~\cite{monras2013phase}):
\begin{align}\label{eq:coherent state QFI, scenario 1}
    \IQ = \frac{4 \bar N \gamma  t_\text{int}^2 (\gamma -\Gamma ) e^{\gamma  t_\text{int}}}{2 \gamma  e^{\gamma  t_\text{int}}-(\gamma +\Gamma ) e^{\Gamma  t_\text{int}}}+\frac{4 \left(\Gamma  e^{\Gamma  t_\text{int}}-e^{\gamma  t_\text{int}} (\Gamma +\gamma  t_\text{int} (\Gamma -\gamma ))\right)^2}{(\gamma -\Gamma )^2 \left(e^{\gamma  t_\text{int}}-e^{\Gamma  t_\text{int}}\right) \left(\gamma  e^{\gamma  t_\text{int}}-\Gamma  e^{\Gamma  t_\text{int}}\right)},
\end{align}
which if we expand in Taylor series in $\gamma/\Gamma\ll1$ yields:
\begin{align}
    \IQ = \frac{4(1- e^{-\Gamma t_\text{int}})}{\Gamma }
    + 4 \gamma  \bar N t_\text{int}^2 e^{-\Gamma t_\text{int}}
    + \frac{4 \gamma  e^{-\Gamma t_\text{int}} (-3 \Gamma  t_\text{int}+3 \sinh (\Gamma  t_\text{int})+\cosh (\Gamma  t_\text{int})-1)}{\Gamma^2 }
    + \mathcal{O}\left[\left(\frac{\gamma}{\Gamma}\right)^2\right],
\end{align}
where the first term is constant, the second term scales as $\gamma\bar N$, and the third term scales as $\gamma$ with no $\bar N$. For $\bar N\gg1$ and $\gamma\ll\Gamma$, we can neglect the third term, leaving just:
\begin{align}\label{eq:coherent qfi}
    \IQ \approx \frac{4(1- e^{-\Gamma t_\text{int}})}{\Gamma }
    + 4 \gamma  \bar N t_\text{int}^2 e^{-\Gamma t_\text{int}}.
\end{align}
The remaining competition between these two terms (i.e.\ the size of $\bar N \gamma \Gamma t_\text{int}^2 e^{-\Gamma t_\text{int}}$ compared to one) comes from the weak signal limit versus the high energy limit, where the time $\Gamma t_\text{int}$ also matters.

In the long dead time limit, we simply optimize the coherent state QFI in Eq.~\ref{eq:coherent qfi} against the interrogation time $t_\text{int}$. This yields $t_{\text{int}}^*=\left[1+\sqrt{1+\Gamma/(\gamma\bar{N})}\right]/\Gamma$, which for $\gamma \bar{N} \gg \Gamma$ converges to $t_{\text{int}}^*\approx2/\Gamma$. In this limit, the optimal coherent state QFI thus converges to $\IQ=\frac{16}{e^{2}}\frac{\gamma\bar{N}}{\Gamma^{2}}+\frac{4(1-1/e^2)}{\Gamma}\approx2.16\frac{\gamma\bar{N}}{\Gamma^{2}}+\frac{3.46}{\Gamma}$. Dropping the second term, since we assume that $\gamma \bar{N} \gg \Gamma$, then the TOP coherent SNR is $\sqrt{\gamma T\IQ/\tau_{\text{dead}}}\approx\sqrt{2.17\bar{N}\gamma^2 T/(\Gamma^{2}\tau_{\text{dead}})}$. This proves the result in the main text.

\clearpage
\newpage
\section{Numerics for independent gain estimation}
\begin{figure}
    \centering
        \includegraphics[width=0.49\columnwidth]{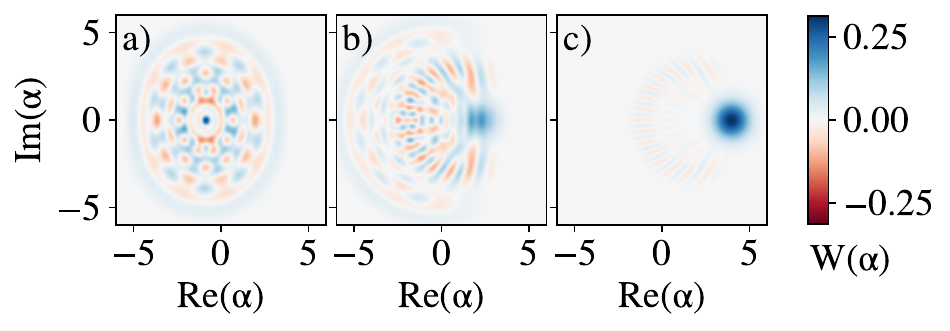}
    \includegraphics[width=0.49\columnwidth]{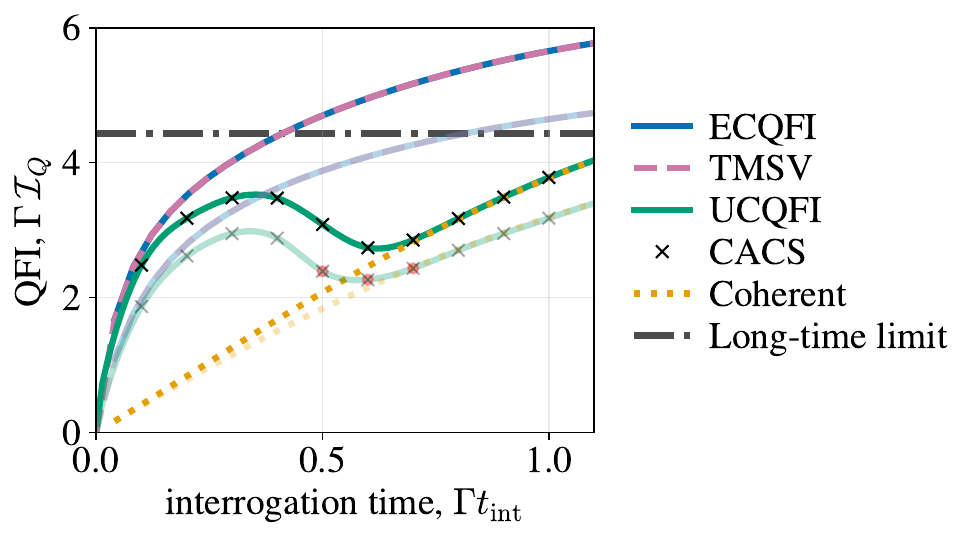}
    \caption{\textbf{(Left)} Wigner functions for the optimal unentangled state from CACS that achieves the UCQFI for $\bar N=8$ (this is a larger scale version of the insets in Fig.~\ref{fig:QFI vs time}). 
    \textbf{(Right)} QFI for estimating $\sqrt\gamma$ versus interrogation time $t_\text{int}$ for different initial states and $\Gamma/\gamma=20$. The results for $\bar N=16$ are shown in opaque lines, whereas those for $\bar N=8$ are shown in partially transparent lines (and are a zoomed-in version of the lines in Fig.~\ref{fig:QFI vs time}). The UCQFI lines are interpolated like in Fig.~\ref{fig:QFI vs time}.
    }
    \label{fig:Wigners and gap SM}
\end{figure}

We now provide some additional details about the numerical calculation of the UCQFI for independent gain estimation of $\sqrt\gamma$ for fixed $\Gamma$.
We use the CACS from Appendix~I of Ref.~\cite{gardner2025stochastic} to calculate the UCQFI numerically. For a given average particle number $\bar N$, we start from either the coherent state or a perturbed coherent state $\ket{\psi}\propto0.9\ket{\alpha}+0.1\ket{\phi}$, where $|\alpha|^2=\bar N$ and $\ket{\phi}$ is chosen Haar-randomly by sampling its Fock basis coefficients from a complex normal distribution and then normalizing. This initial perturbed state may not satisfy the average particle number constraint $\bar N$, but the algorithm ensures that all subsequent states will. Our results are similar if we start from a fully random state rather than the coherent state, but we observe that the QFI at long times converges to the coherent state faster if we start from a perturbed coherent state.
The CACS results in Fig.~\ref{fig:QFI vs time} for $\bar N=8$ are after 1000 iterations starting from the coherent state in a truncated Hilbert space of dimension 20.  
The Wigner function plots for $\bar N=8$ shown in the insets of Fig.~\ref{fig:QFI vs time} are shown on a larger scale in Fig.~\ref{fig:Wigners and gap SM}: these show the non-Gaussian to Gaussian transition in the optimal unentangled state at the non-monotonic dip in the UCQFI.
We check that the entanglement gap is a property of the quantum channel by increasing the average particle number from $\bar N=8$ to $\bar N=16$. We observe that the entanglement gap does not close and the non-monotonic dip does not move significantly to later times, as shown in Fig.~\ref{fig:Wigners and gap SM}. 
The CACS results in Fig.~\ref{fig:Wigners and gap SM} for $\bar N=16$ are for 100 iterations starting from a perturbed coherent state in a truncated Hilbert space of dimension 40. Although biconvex, the CACS is still computationally expensive, which is why we interpolate the UCQFI curve. We see that Fock states are optimal at short times, which is why we use them there as part of the interpolated spline. We use a spline that follows the Fock results below $\Gamma t_\text{int}=0.04$, then the ten CACS data points, and then the coherent results above $\Gamma t_\text{int}=1.1$. At long times, it appears that coherent states are optimal, from extrapolating the CACS results, so we use them there as part of the interpolated spline. With more computing resources, future work could sample a larger and denser region of parameter space using the CACS and avoid the interpolated spline, which leads to some roughness in Fig.~\ref{fig:dead_time}. The smoother spline still looks better and is more representative of the underlying physics than, e.g., taking a direct linear interpolation of the data.

Similarly to the \textit{Applications} section in the main text, which considers noise sensing, we can also plug in some numbers for independent gain estimation, as shown in Table~\ref{tab:TOP SNR numerics SM}. Recall that we have that $\bar N\approx 5$, $1/\Gamma \approx 100\,\mu$s, and $\tau_{\rm dead}\approx 200\,\mu$s. The gain in SNR from using entanglement is $15\%$ compared to the coherent state.
If we were able to increase the energy to $\bar N=8$, then the gain from using entanglement is instead $11\%$ compared to the UCQFI (or $17\%$ compared to the coherent state).
This indicates that a resonator may be prepared in an entangled state with an ancilla for incremental SNR gain.

\setlength{\tabcolsep}{6pt}
\begin{table}\begin{tabular}{@{}lll@{}}
\toprule
Average number                            & $\bar N=5$ & $\bar N=8$ \\ \midrule
ECQFI & $0.027\; [\Gamma t_\text{int}^*=0.76]$ & $0.125\; [\Gamma t_\text{int}^*=0.71]$ \\
Coherent state & $0.023\; [\Gamma t_\text{int}^*=1.61]$ & $0.108\; [\Gamma t_\text{int}^*=1.61]$ \\
UCQFI & (not computed) & $0.113\; [\Gamma t_\text{int}^*=0.31]$ \\ \bottomrule
\end{tabular}\caption{Time-optimized (TOP) SNR in units of $\sqrt{T}$ for different strategies and energies. The optimal interrogation time $t_\text{int}^*$ is shown in brackets $[\cdot]$. We assume that $\Gamma \tau_\text{dead}=2$ and $\gamma/\Gamma=0.05$.
}
\label{tab:TOP SNR numerics SM}
\end{table}

\newpage

\section{Long interrogation time limit}
Let us now find the optimal state in the steady state limit of long interrogation time $t\to\infty$. This corresponds to $\eta\to0$ so it can also be thought of as the low transmissivity limit. We will show that our main result in Claim~\ref{claim:ECQFI and TMSV} for a generic parameter of the thermal loss channel can be independently shown using a proof similar to the low-transmissivity quantum illumination result from Ref.~\cite{PhysRevLett.118.070803}.

We start with the $\eta\to0$ limit of our main result. In this $\eta\to0$ limit, the coefficients in Eq.~\ref{eq:coeffs, EM} become:
\begin{align*}     c_1 &= \bar n_{\text{th}} (\bar n_{\text{th}}+1)\dot{\eta }^2|_{\eta=0},  \\
    c_2 &= \bar n_{\text{th}} (\bar n_{\text{th}}+1) \dot{\eta }^2|_{\eta=0} 
        - 4 \eta  \bar n_{\text{th}} (\bar n_{\text{th}}+1) \dot{\eta }|_{\eta=0} \dot{\bar n}_{\text{th}}|_{\eta=0}
        + \eta  (2 \bar n_{\text{th}}+1) \dot{\bar n}_{\text{th}}^2|_{\eta=0}, \nonumber\\
    c_3 &= \eta  (\bar n_{\text{th}}+1) [\bar n_{\text{th}}\dot{\eta } -\dot{\bar n}_{\text{th}}]^2, \nonumber\\
    c_4 &= \eta \bar n_{\text{th}} (\bar n_{\text{th}}+1) (2 \bar n_{\text{th}}+1), \nonumber\\
    c_5 &= \eta \bar n_{\text{th}} (\bar n_{\text{th}}+1)^2,\nonumber
\end{align*}
where we have to be careful not to take $\eta=0$ of terms involving $\dot\eta$ yet without accounting for $\dot\eta^2=4\eta\dot\zeta^2$ with the amplitude transmissivity $\zeta:=\sqrt{\eta}$ such that:
\begin{align*}     c_1 &= \bar n_{\text{th}} (\bar n_{\text{th}}+1)4\eta\dot\zeta^2|_{\eta=0},  \\
    c_2 &= \bar n_{\text{th}} (\bar n_{\text{th}}+1) 4\eta\dot\zeta^2|_{\eta=0}
        - 4 \eta  \bar n_{\text{th}} (\bar n_{\text{th}}+1) 2\sqrt{\eta}\dot\zeta|_{\eta=0} \dot{\bar n}_{\text{th}}|_{\eta=0}
       + \eta  (2 \bar n_{\text{th}}+1) \dot{\bar n}_{\text{th}}^2|_{\eta=0}, \nonumber\\
    c_3 &= \eta  (\bar n_{\text{th}}+1) [\bar n_{\text{th}}2\sqrt\eta\dot\zeta|_{\eta=0} -\dot{\bar n}_{\text{th}}]^2, \nonumber\\
    c_4 &= \eta \bar n_{\text{th}} (\bar n_{\text{th}}+1) (2 \bar n_{\text{th}}+1), \nonumber\\
    c_5 &= \eta \bar n_{\text{th}} (\bar n_{\text{th}}+1)^2,\nonumber
\end{align*}
and the ECQFI in Eq.~\ref{eq:ECQFI} becomes, dropping $\mathcal{O}(\eta^{3/2})$ terms in the numerator since they yield $\mathcal{O}(\sqrt\eta)$ terms after division:
\begin{align*}
    \IQ^\text{E}
    &=\frac{\bar n_{\text{th}} (\bar n_{\text{th}}+1)4\eta\dot\zeta^2|_{\eta=0}\bar N^2+[\bar n_{\text{th}} (\bar n_{\text{th}}+1) 4\eta\dot\zeta^2|_{\eta=0}+ \eta  (2 \bar n_{\text{th}}+1) \dot{\bar n}_{\text{th}}^2|_{\eta=0}]\bar N+\eta  (\bar n_{\text{th}}+1) \dot{\bar n}_{\text{th}}^2}{\eta \bar n_{\text{th}} (\bar n_{\text{th}}+1) (2 \bar n_{\text{th}}+1)\bar N+\eta \bar n_{\text{th}} (\bar n_{\text{th}}+1)^2}
    \\&=\frac{\bar n_{\text{th}} (\bar n_{\text{th}}+1)4\bar N^2+\bar n_{\text{th}} (\bar n_{\text{th}}+1) 4\bar N}{\bar n_{\text{th}} (\bar n_{\text{th}}+1) (2 \bar n_{\text{th}}+1)\bar N+\bar n_{\text{th}} (\bar n_{\text{th}}+1)^2} \dot\zeta^2|_{\eta=0}
    +\frac{(2 \bar n_{\text{th}}+1) \bar N+ (\bar n_{\text{th}}+1) }{\bar n_{\text{th}} (\bar n_{\text{th}}+1) (2 \bar n_{\text{th}}+1)\bar N+\bar n_{\text{th}} (\bar n_{\text{th}}+1)^2} \dot{\bar n}_{\text{th}}^2|_{\eta=0}
    \\&=\frac{4\bar N (\bar N+1)}{(2 \bar n_{\text{th}}+1)\bar N+\bar n_{\text{th}}+1} \dot\zeta^2|_{\eta=0}
    +\frac{1}{\bar n_{\text{th}} (\bar n_{\text{th}}+1)} \dot{\bar n}_{\text{th}}^2|_{\eta=0},
\end{align*}
which becomes after change of variable back to $\eta$:
\begin{align}\label{eq:long time limit}
    \IQ^\text{E}
    =\frac{4\bar N (\bar N+1)}{(2 \bar n_{\text{th}}+1)\bar N+\bar n_{\text{th}}+1} \frac{\dot\eta^2|_{\eta=0}}{4\eta}
    +\frac{1}{\bar n_{\text{th}} (\bar n_{\text{th}}+1)} \dot{\bar n}_{\text{th}}^2|_{\eta=0},
\end{align}
where the $1/(4\eta)$ is just a coordinate singularity from using power transmissivity $\eta$ instead of amplitude transmissivity $\zeta=\sqrt\eta$.

We now show that we can recover this result for $\eta\to0$ independently from our main proof. We closely follow the proof in Ref.~\cite{PhysRevLett.118.070803}, but now for a generic parameter $\theta$. We again define $\zeta:=\sqrt\eta$ as the amplitude transmissivity.
The initial state is $\ket{\Psi}=\sum_\alpha \sqrt{p_\alpha}\ket{w_\alpha}\otimes\ket{v_\alpha}$ and the final state is $\rho(\theta)=\text{Tr}_S[U_\zeta\proj{\Psi}\otimes\rho_B U_\zeta^\dagger]$. For $\zeta\to0$, i.e.\ $t\to\infty$, then $\rho(\theta)|_{\zeta=0}=\sum_\alpha p_\alpha \proj{v_\alpha}\otimes\rho_B$, where $U_\zeta\approx\exp(\zeta[s^\dagger b - s b^\dagger])$ and the thermal steady state is $\rho_B=\sum_n \rho_n \proj{n}$ with $\rho_n= \frac{\bar n_\text{th}^n}{(1+\bar n_\text{th})^{n+1}} $. The final state $\rho(\theta)$ is diagonal in $\ket{\phi_{\alpha n}}=\ket{v_\alpha}\otimes\ket{n}$ with $\lambda_{\alpha n}=p_\alpha\rho_n$. Its derivative is the following:
\begin{align*}
    \partial_{\theta}\rho(\theta)|_{\zeta=0}
    &= \partial_{\theta}\text{Tr}_S[U_\zeta\proj{\Psi}\otimes\rho_B U_\zeta^\dagger]|_{\zeta=0}
    \\&= \text{Tr}_S[(\deriv{\zeta}{\theta}\partial_{\zeta} U_\zeta)\proj{\Psi}\otimes\rho_B U_\zeta^\dagger]|_{\zeta=0} + \text{Tr}_S[U_\zeta\proj{\Psi}\otimes\rho_B (\deriv{\zeta}{\theta}\partial_{\zeta} U_\zeta)^\dagger]|_{\zeta=0} + \text{Tr}_S[U_\zeta\proj{\Psi}\otimes(\deriv{\bar n_\text{th}}{\theta}\partial_{\bar n_\text{th}} \rho_B) U_\zeta^\dagger]|_{\zeta=0}
        \\&= \deriv{\zeta}{\theta}|_{\zeta=0}\text{Tr}_S[[s^\dagger b - s b^\dagger,\proj{\Psi}\otimes\rho_B]] + \deriv{\bar n_\text{th}}{\theta} \sum_\alpha p_\alpha \proj{v_\alpha}\otimes\partial_{\bar n_\text{th}} \rho_B 
\end{align*}
From Eq.~2 of the Supplemental Material of Ref.~\cite{PhysRevLett.118.070803}:
\begin{align*}
    \text{Tr}_S[[s^\dagger b - s b^\dagger,\proj{\Psi}\otimes\rho_B]]
    &= \sum_{\alpha\alpha'} \sqrt{p_\alpha p_{\alpha'}}\ket{v_\alpha}\bra{v_{\alpha'}}\otimes[\braopket{w_{\alpha'}}{s^\dagger}{w_\alpha}b - \braopket{w_{\alpha'}}{s}{w_\alpha}b^\dagger,\rho_B]
    .
\end{align*}
The derivative of the steady state itself is:
\begin{align*}
    \partial_{\bar n_\text{th}} \rho_B 
    = \partial_{\bar n_\text{th}}\sum_n \rho_n \proj{n}
    = \partial_{\bar n_\text{th}}\sum_n  \frac{\left(\bar n_\text{th}\right)^{n}  }{\left(1+\bar n_\text{th}\right)^{n+1}  }\proj{n} 
    = \frac{1}{\left(1+\bar n_\text{th}\right)^{2}  } \sum_n  \left(n-\bar n_\text{th}\right) \rho_{n-1} \proj{n} ,
\end{align*}
where $\rho_{-1}=1/\bar n_\text{th}$.
The full derivative is thus:
\begin{align*}
    \partial_{\theta}\rho(\theta)|_{\zeta=0}
    &= \deriv{\zeta}{\theta}|_{\zeta=0}\sum_{\alpha\alpha'} \sqrt{p_\alpha p_{\alpha'}}\ket{v_\alpha}\bra{v_{\alpha'}}\otimes[\braopket{w_{\alpha'}}{s^\dagger}{w_\alpha}b - \braopket{w_{\alpha'}}{s}{w_\alpha}b^\dagger,\rho_B ]
    + \deriv{\bar n_\text{th}}{\theta}\frac{1}{(1+\bar n_\text{th})^2}  \sum_\alpha p_\alpha \proj{v_\alpha}\otimes\sum_n \left(n-\bar n_\text{th}\right) \rho_{n-1} \proj{n}
    \\&=: \dot{\zeta} X + \dot{\bar n}_\text{th} Y,
\end{align*}
where $X$ and $Y$ are defined implicitly.
The QFI with respect to $\theta$ can now be calculated similarly to Ref.~\cite{PhysRevLett.118.070803}:
\begin{align*}
    \IQ|_{\zeta\to0} 
    &= \sum_{\alpha\alpha'nn'} \frac{2}{\lambda_{\alpha n}+\lambda_{\alpha' n'}}\abs{\braopket{\phi_{\alpha n}}{\partial_{\theta}\rho(\theta)|_{\zeta\to0}}{\phi_{\alpha' n'}}}^2
    \\&= \sum_{\alpha\alpha'nn'} \frac{2}{\lambda_{\alpha n}+\lambda_{\alpha' n'}}\abs{\braopket{\phi_{\alpha n}}{\dot{\zeta} X + \dot{\bar n}_\text{th}  Y}{\phi_{\alpha' n'}}}^2
    \\&= \dot{\zeta}^2 H + \dot{\bar n}_\text{th} ^2 \sum_{\alpha\alpha'nn'} \frac{2}{\lambda_{\alpha n}+\lambda_{\alpha' n'}}\abs{\braopket{\phi_{\alpha n}}{Y}{\phi_{\alpha' n'}}}^2 
        \\&+ \dot{\zeta} \dot{\bar n}_\text{th}  \sum_{\alpha\alpha'nn'} \frac{2}{\lambda_{\alpha n}+\lambda_{\alpha' n'}}(\braopket{\phi_{\alpha n}}{X}{\phi_{\alpha' n'}}\braopket{\phi_{\alpha n}}{Y}{\phi_{\alpha' n'}}^*+\braopket{\phi_{\alpha n}}{Y}{\phi_{\alpha' n'}}\braopket{\phi_{\alpha n}}{X}{\phi_{\alpha' n'}}^*)
    \\&= \dot{\zeta}^2 \mathfrak H + \dot{\bar n}_\text{th} ^2 \frac{1}{\bar n_\text{th}(1+\bar n_\text{th})} + \dot{\zeta} \dot{\bar n}_\text{th} Z
    ,
\end{align*}
where $\mathfrak H:=\frac{4}{1+\bar n_\text{th}}\sum_{\alpha\alpha'}\frac{p_\alpha p_{\alpha'}}{p_{\alpha'}+p_\alpha\frac{\bar n_\text{th}}{1+\bar n_\text{th}}}\abs{\braopket{w_{\alpha'}}{s}{w_{\alpha}}}^2$ is the QFI in Eq.~3 of Ref.~\cite{PhysRevLett.118.070803} with respect to $\zeta$.
The $\dot{\bar n}_\text{th} ^2$ term is the QFI of the thermal steady state $\rho_B$ with respect to $\theta$. The QFI with respect to $\bar n_\text{th}=\frac{\gamma}{\Gamma-\gamma}$ is $\IQ=\frac{1}{\bar n_\text{th}(1+\bar n_\text{th})}$. Let us calculate the $\dot{\zeta} \dot{\bar n}_\text{th} Z $ cross term. Ref.~\cite{PhysRevLett.118.070803} shows that:
\begin{align*}
    \braopket{\phi_{\alpha n}}{X}{\phi_{\alpha' n'}}
            &= \sqrt{p_\alpha p_{\alpha'}}(\rho_{n'}-\rho_n)\left(\braopket{w_{\alpha'}}{s^\dagger}{w_\alpha}\sqrt{n+1}\delta_{n',n+1}-\braopket{w_{\alpha'}}{s}{w_\alpha}\sqrt{n'+1}\delta_{n,n'+1}\right)
\end{align*}
and we have that:
\begin{align*}
    \braopket{\phi_{\alpha n}}{Y}{\phi_{\alpha' n'}}
    &= \frac{1}{(1+\bar n_\text{th})^2} \bra{v_\alpha}\otimes\bra{n} \sum_\beta p_\beta \proj{v_\beta}\otimes\sum_m (m-\bar n_\text{th}) \rho_{m-1} \proj{m} \ket{v_{\alpha'}}\otimes\ket{n'} 
    = \frac{1}{(1+\bar n_\text{th})^2} \delta_{\alpha\alpha'}p_\alpha \delta_{nn'}  (n-\bar n_\text{th}) \rho_{n-1}
\end{align*}
which is real such that $\braopket{\phi_{\alpha n}}{Y}{\phi_{\alpha' n'}}^* $ takes the same value. Therefore, we have the following plus its complex conjugate for the $\dot{\zeta} \dot{\bar n}_\text{th} Z $ cross term:
\begin{align*}
    &\sum_{\alpha\alpha'nn'} \frac{2}{\lambda_{\alpha n}+\lambda_{\alpha' n'}} \braopket{\phi_{\alpha n}}{X}{\phi_{\alpha' n'}}\braopket{\phi_{\alpha n}}{Y}{\phi_{\alpha' n'}}^*
    \\&= \sqrt{p_\alpha p_{\alpha'}}(\rho_{n'}-\rho_n)\left(\braopket{w_{\alpha'}}{s^\dagger}{w_\alpha}\sqrt{n+1}\delta_{n',n+1}-\braopket{w_{\alpha'}}{s}{w_\alpha}\sqrt{n'+1}\delta_{n,n'+1}\right) \frac{1}{(1+\bar n_\text{th})^2} \delta_{\alpha\alpha'}p_\alpha \delta_{nn'}  (n-\bar n_\text{th}) \rho_{n-1}
    \\&= 0
\end{align*}
such that the cross term vanishes in the QFI above which becomes:
\begin{align*}
    \IQ|_{\zeta\to0} 
            =\dot{\eta}^2 \frac{\mathfrak H}{4\eta} + \dot{\bar n}_\text{th} ^2 \frac{1}{\bar n_\text{th}(1+\bar n_\text{th})},
\end{align*}
where the $1/(4\eta)$ factor comes from change of variable from $\zeta$ to $\eta$. 

We now have to show that $\mathfrak H$ is maximized by TMSV for a given $\bar N$ constraint.
Let $q:=\frac{\bar n_{\rm th}}{1+\bar n_{\rm th}}$ with $q>0$. First, we notice that, for any $x,y\geq0$ and any $\lambda>0$, the following inequality holds:
\begin{align*}
    \frac{xy}{y+qx}
    &\leq
    \frac{x+q\lambda^2 y}{(1+q\lambda)^2},
\end{align*}
since this is equivalent to:
\begin{align*}
    (x+q\lambda^2 y)(y+qx)
    -(1+q\lambda)^2xy
    =
    q(x-\lambda y)^2
    \geq 0 .
\end{align*}
Applying this inequality with $x=p_\alpha\geq0$ and $y=p_{\alpha'}\geq0$ and $q=\frac{\bar n_{\rm th}}{1+\bar n_{\rm th}}$ for arbitrary $\lambda$ gives:
\begin{align*}
    \mathfrak H
    =\frac{4}{1+\bar n_\text{th}}\sum_{\alpha\alpha'}\frac{p_\alpha p_{\alpha'}}{p_{\alpha'}+p_\alpha\frac{\bar n_\text{th}}{1+\bar n_\text{th}}}\abs{\braopket{w_{\alpha'}}{s}{w_{\alpha}}}^2
    &\leq
    \frac{4}{1+\bar n_{\rm th}}
    \frac{1}{(1+q\lambda)^2}
    \sum_{\alpha\alpha'}
    \left(
        p_\alpha+q\lambda^2 p_{\alpha'}
    \right)
    \abs{\braopket{w_{\alpha'}}{s}{w_\alpha}}^2 .
\end{align*}
Using completeness of the Schmidt basis and $[s,s^\dagger]=1$, we have that:
\begin{align*}
    \sum_{\alpha\alpha'}
    p_\alpha
    \abs{\braopket{w_{\alpha'}}{s}{w_\alpha}}^2
    &=
    \sum_\alpha p_\alpha
    \braopket{w_\alpha}{s^\dagger s}{w_\alpha}
    =
    \bar N,
    \\
    \sum_{\alpha\alpha'}
    p_{\alpha'}
    \abs{\braopket{w_{\alpha'}}{s}{w_\alpha}}^2
    &=
    \sum_{\alpha'} p_{\alpha'}
    \braopket{w_{\alpha'}}{s s^\dagger}{w_{\alpha'}}
    =
    \bar N+1.
\end{align*}
Inserting these into the expression above yields the following inequality for a given $\lambda$:
\begin{align*}
    \mathfrak H
    &\leq
    \frac{4}{1+\bar n_{\rm th}}
    \frac{\bar N+q\lambda^2(\bar N+1)}
    {(1+q\lambda)^2}.
\end{align*}
This is minimized at $\lambda=\frac{\bar N}{\bar N+1}$ for $\bar n_\text{th}>0$, which gives:
\begin{align*}
    \mathfrak H
    &\leq
    \frac{4}{1+\bar n_{\rm th}}
    \frac{\bar N(\bar N+1)}
    {(1+q)\bar N+1}
    =
    \frac{4\bar N(\bar N+1)}
    {(2\bar n_{\rm th}+1)\bar N+\bar n_{\rm th}+1}.
\end{align*}
This upper bound is saturated by TMSV by comparison to the TMSV QFI in Eq.~6 of Ref.~\cite{PhysRevLett.118.070803}. We emphasize that, compared to that work, our result holds for any $\bar N$ and any parameter of the thermal loss channel. In fact, since the TMSV state is:
\begin{align*}
    \ket{{\rm TMSV}}
    =
    \sum_{n=0}^{\infty}
    \sqrt{(1-\lambda)\lambda^n}\ket n_I\ket n_S,
    \qquad
    \lambda=\frac{\bar N}{\bar N+1},
\end{align*}
where $p_n=(1-\lambda)\lambda^n=\lambda p_{n-1}$, then we can understand that this saturates the inequality because $x=\lambda y$ for $x=p_\alpha\geq0$ and $y=p_{\alpha'}\geq0$ since only neighboring ladder terms appear in $\abs{\braopket{w_{\alpha'}}{s}{w_{\alpha}}}^2=\abs{\braopket{n'}{s}{n}}^2=\abs{\braopket{n'}{\sqrt{n}}{n-1}}^2=n\delta_{n',n-1}$. Since TMSV attains the bound, it therefore maximizes $\mathfrak H$. Substitution of $\mathfrak H=\frac{4\bar N(\bar N+1)}{(2\bar n_{\rm th}+1)\bar N+\bar n_{\rm th}+1}$ recovers the QFI in Eq.~\ref{eq:long time limit}, as required.

Therefore, since the $\dot {\bar n}_\text{th}$ term is state independent, the same conclusions as Ref.~\cite{PhysRevLett.118.070803} about the entanglement gap in the $\zeta\to0$ limit apply if the choice of parameter $\theta$ to estimate has $\dot{\zeta}\neq0$: The coherent state is the optimal unextended state and there is a finite gap per photon to the optimal extended state which is TMSV. This gap vanishes in the limit of $\bar N\to\infty$. 
The exact value of the gap is modified by the steady state term and the choice of parameter $\theta$.

If instead $\dot{\zeta}=0$, e.g.\ for $\theta=\sqrt\gamma$ where $\dot\zeta\to0$ as $t\to\infty$, then only the steady-state term remains and there is no entanglement gap, as shown in Fig.~\ref{fig:QFI vs time}. This is because $\dot\zeta^2=\dot\eta^2/(4\eta)=\gamma t^2\eta\to0$ as $t\to\infty$. Then, we have that the QFI with respect to $\sqrt\gamma$ from the steady state term is $\IQ=\frac{4\Gamma}{(\Gamma-\gamma)^2}$ which recovers the familiar $4/\Gamma$ for $\gamma\ll\Gamma$.

\end{document}